# Direct visualization of ultrafast lattice ordering triggered by an electron-hole plasma in 2D perovskites

Hao Zhang[1,2], Wenbin Li[1,2], Joseph Essman[1], Claudio Quarti[3,4], Isaac Metcalf[5], Wei-Yi Chiang[6], Siraj Sidhik[1,5], Jin Hou[5], Austin Fehr[1], Andrew Attar[7], Ming-Fu Lin[7], Alexander Britz[7], Xiaozhe Shen[7], Stephan Link[6], Xijie Wang[7], Uwe Bergmann[8,9], Mercouri G. Kanatzidis[10], Claudine Katan[3], Jacky Even[11], Jean-Christophe Blancon[1]* and Aditya D. Mohite[1]*

**[1]Department of Chemical and Biomolecular Engineering, Rice University, Houston, Texas 77005, USA.**

**[2]Applied Physics Program, Smalley-Curl Institute, Rice University, Houston, TX, 77005, USA.**

**[3]Univ Rennes, ENSCR, INSA Rennes, CNRS, ISCR (Institut des Sciences Chimiques de Rennes) - UMR 6226, F-35000 Rennes, France.**

**[4]Laboratory for Chemistry of Novel Materials, Department of Chemistry, University of Mons, Place du Parc 20, 7000 Mons, Belgium**

**[5]Department of Materials Science and NanoEngineering, Rice University, Houston, TX, 77005, USA.**

**[6]Department of Chemistry, Rice University, Houston, Texas 77005, USA.**

**[7]SLAC National Accelerator Laboratory, Menlo Park, CA 94025, USA.**

**[8]PULSE Institute, SLAC National Accelerator Laboratory, Stanford University, Stanford, CA, USA**

**[9]Department of Physics, University of Wisconsin–Madison, Madison, WI 53706 USA.**

**[10]Department of Chemistry, Northwestern University, Evanston, Illinois 60208, USA.**

**[11]Univ Rennes, CNRS, Institut FOTON (Fonctions Optiques pour les Technologies de l'Information), UMR 6082, CNRS, INSA de Rennes, 35708 Rennes, France.**

***Correspondence: blanconjc@gmail.com and adm4@rice.edu**

**Direct visualization of ultrafast coupling between charge carriers and lattice degrees of freedom in photo-excited semiconductors has remained a long-standing challenge and is critical for understanding the light-induced physical behavior of materials under extreme non-equilibrium conditions. Here, by monitoring the evolution of the wave-vector resolved ultrafast electron diffraction intensity following above-bandgap photo-excitation, we obtain a direct visual of the structural dynamics in monocrystalline 2D perovskites. Analysis reveals a surprising, light-induced ultrafast lattice ordering resulting from a strong interaction between hot-carriers and the perovskite lattice, which induces an in-plane octahedra rotation, towards a more symmetric phase. Correlated ultrafast spectroscopy performed at the same**

**carrier density as ultrafast electron diffraction reveals that the creation of a hot and dense electron-hole plasma triggers lattice ordering at short timescales by modulating the crystal cohesive energy. Finally, we show that the interaction between the carrier gas and the lattice can be altered by tailoring the rigidity of the 2D perovskite by choosing the appropriate organic spacer layer.**

Organic-inorganic (hybrid) two-dimensional (2D) halide perovskites are constructed by a superlattice of interlocking organic and inorganic nanometer-thick layers and have demonstrated unique and non-classical behaviors and are being extensively explored for a wide range of technologies such as photovoltaics, photodetectors, photocatalysts, light emitting diodes, lasers, and quantum emitters.[1–9] The underlying design principles for each of these devices are strongly correlated to the exact details of how photoexcited or electronically injected charge carriers dissipate their energy via electron-phonon coupling. For example, it has been shown recently, that unusual electron-phonon coupling mechanisms are likely to promote emission of single photons or correlated photons pairs from perovskite quantum sources.[10] There have been only handful experimental studies, which have employed ultrafast or temperature dependent optical spectroscopy to elucidate carrier dynamics in 2D perovskites. Room temperature coherent oscillations were observed from two-dimensional coherent excitation spectroscopy in 2D perovskites with aromatic barriers as organic barriers.[11] These oscillations were attributed to a combination of a resonant effect within the exciton fine structure and to electron-phonon coupling with low–frequency optical modes. Recent work by Quan et al used optical pump-probe spectroscopy to show that vibrational relaxation dynamics in 2D perovskites formed from flexible alkyl-amines as organic barriers is fast and relatively independent of the lattice temperature, whereas  from aromatic amines is slower, and are temperature dependent.[12] Moreover, Gong et al has shown the crystal rigidity strongly affects the strength of electron-phonon coupling and therefore promotes carrier trapping, leading to a fast non-radiative decay.[5] These measurements show that presence of an organic cation in close proximity to the inorganic perovskite lattice strongly modulates the nature of electron-phonon interactions,[5,11–18] and suggested that electron-phonon scattering in 2D perovskites occur via, local dynamic disorder.[6] These short-range carrier-lattice interactions modulates the quantum-well thickness and octahedral tilts, leading to exciton self-trapping and broadband emission, thanks to the lattice softness, suggesting potential benefit of 2D perovskites in white-light applications. However, there exists no direct ultrafast structural measurement of 2D perovskites, which elucidates the underlying carrier-lattice interaction mechanisms upon optical excitation. Moreover, the exact geometry and dynamics of these lattice distortions are unknown. This is largely due to the challenges in visualizing the carrier-lattice coupling and dynamics that result after optical excitation above the band-gap.

Here we report the first direct measurement of structural dynamics in 2D perovskites obtained by monitoring the change in the ultrafast femtosecond electron diffraction (UED) after optical excitation. This technique enables a time-resolved structural evolution of the 2D perovskites by tracking the changes in the diffraction pattern, thus providing a direct visualization of lattice response during carrier cooling. Detailed analysis of the Bragg peak intensities and temporal signatures reveals a non-classical light-induced lattice ordering at early times (≤1 ps), which is attributed to the light-induced in-plane rotation of the perovskite octahedra from a distorted toward a symmetrical phase. Complementary transient absorption measurements further reveal a high excitation regime (beyond Mott transition),[19,20] where hot and dense electron-hole plasma strongly modulates the crystal cohesive energy,[21] leading to an ultrafast lattice ordering. In parallel, a classical energy transfer to the whole phonon bath via thermal atomic displacements was observed, which was attributed to the Debye-Waller effect. The Debye-Waller effect was associated with a slower rise time (~ 5ps) of the thermal dissipation (or lattice heating) leading to coherent acoustic oscillations over longer time. Concomitant with the Debye-Waller effect, we also observe an increase in the diffused scattering, which confirms the activation of thermal transfer to the phonon bath. Finally, we show that the mechanism and dynamics of the interaction between the charge carriers and the lattice are acutely tunable and sensitive to the rigidity of the 2D perovskites dictated by the choice of the organic spacer layer. These findings reveal new and distinct carrier-lattice interactions and counterintuitive mechanisms in 2D perovskites, which have not been reported in classical materials such as Si, GaAs or even in 3D halide perovskites.[22,23]

## UED experiment on 2D perovskites

The UED experiments are performed in a pump-probe setup in which sub-micron thick 2D hybrid perovskite single crystals are photoexcited with a 75-fs pulse laser and the structural dynamics are probed with a pulsed electron beam with 150 fs temporal width after specific delay times (t) (Fig.

1a). A diffraction image is acquired at each delay time after above-gap excitation (Fig. 1b), which reflects the instantaneous 2D perovskite crystal structure. Monitoring the characteristics of the diffraction peaks (or Bragg peaks) on the image allows us to directly probe the dynamic changes in the lattice structure after light excitation. These changes directly reflect the carrier-lattice interactions during energy relaxation (cooling) of the high-energy photoexcited carriers (hot carriers) to the 2D perovskite semiconductors band-edge. We start with a phase-pure (homogenous perovskite layer thickness) 2D perovskite crystal of Dion-Jacobson 4AMP-$MAPb_2I_7$ (DJ n=2) with thickness of 270 nm (Fig. 1a and Fig. S1), which is excited with 3.1 eV light, i.e. 0.9 eV above its ground state optical transition[24]. The diffraction pattern of DJ n=2 at rest is consistent with the static crystal structure reported previously (Fig. 1c and 1d, and details in Fig. S2),[24] which exhibits antiferro-distortions (clockwise and anticlockwise octahedral rotations around the **c** axis) associated with an in-plane doubling of the unit cell (in-plane is defined by **a** and **b** axes).[25,26] The diffraction pattern also indicates that the orientation of the crystal layers is parallel to the substrate (see Supplementary discussion 1 and Fig. S2), with both the light excitation and the probe electron beams impinging along the DJ n=2 stacking axis (**c** axis), as depicted in Fig. 1a.

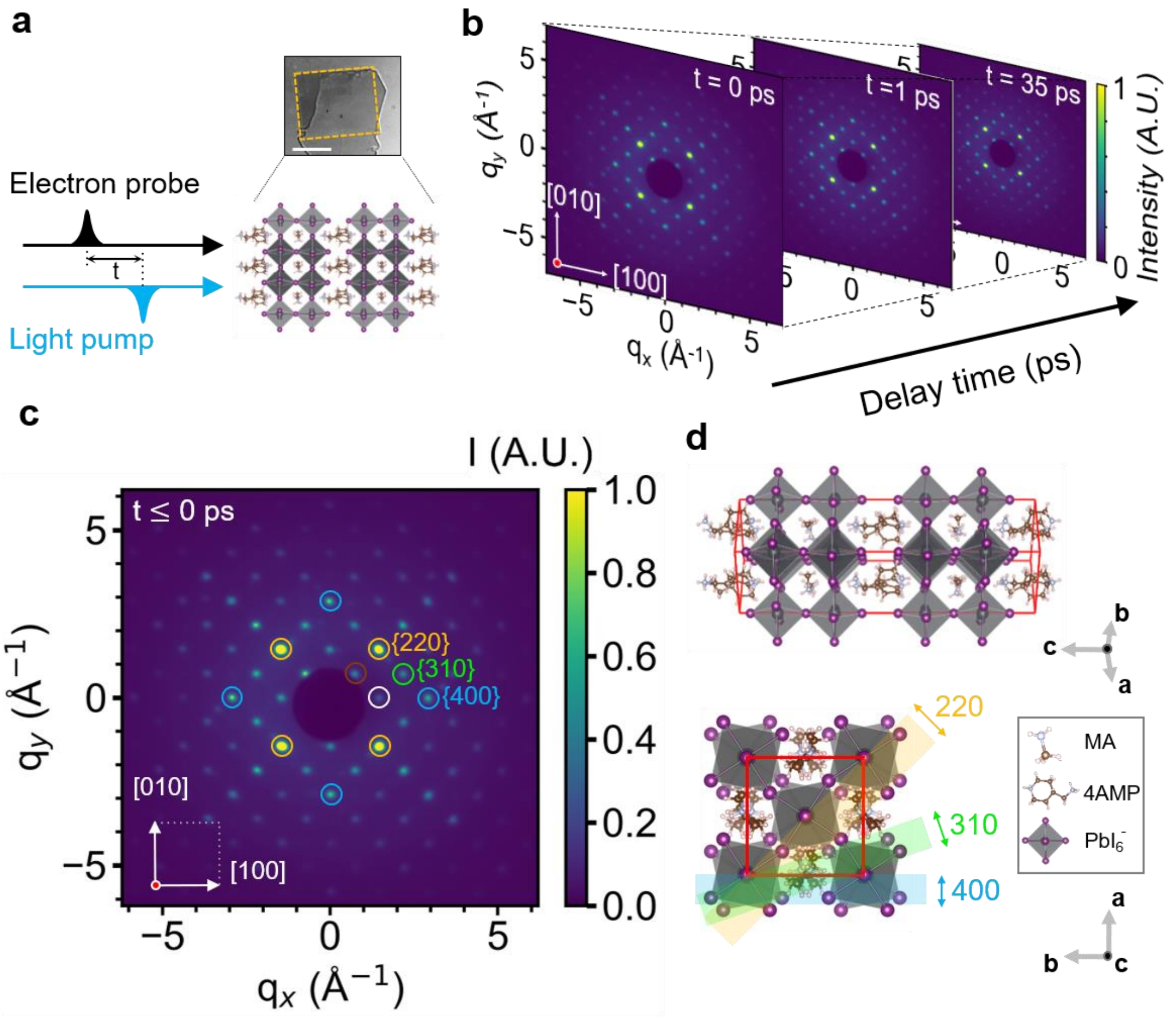

**Fig. 1 | UED measurements on 4AMP-$MAPb_2I_7$ (DJ n=2) 2D perovskite. a,** Schematics of UED apparatus. The directions of light pump and electron probe with respect to crystal orientation is illustrated. (inset) Optical microscopic top-view image of DJ n=2 single crystal. The dashed yellow line idicates the transmission window for the electron beam. Scale bar is 50µm. **b,** Electron diffraction pattern of DJ n=2 crystal at different pump-probe delay time: t = 0 ps, 1 ps and 35 ps. **c,** Static diffraction pattern (t≤0) and indexing of Bragg planes (hk0). Diffraction peaks circled with white, brown, cyan, green, and yellow colors belong to {200}, {110}, {400}, {310}, and {220} Bragg peak families, respectively. The diffraction peaks {110} are partialled blocked by the direct beam filter ($q<1Å^{-1}$). **d,** Crystal structure of DJ n=2 viewing at [110] and [001] directions, showing intrinsic antiferro-distortion (in-plane octahedral rotations). The corresponding Bragg plane directions are displayed for {400} (cyan),{310} (green) and {220} (yellow).

Fig. 2a shows the differential diffraction map indicating the change in the Bragg peak intensities of the DJ n=2 crystal that occurs within a few picoseconds (t =2 ps, averaged within 1~3ps) after light excitation with a fluence of 2 $mJ/cm^2$ (corresponds to a carrier density of $2.5\times10^{13} cm^{-2}$, see Supplementary discussion 2). In this map, the blue and red spots correspond, respectively, to a relative decrease and increase in the peak intensities after photoexcitation with respect to the sample at rest. A first visual inspection of this data reveals both an anisotropic response with respect to the in-plane directions in the 2D perovskite lattice and a monotonically decreasing of absolute intensity response with respect to the magnitude of the scattering vector |q|. The Bragg peaks that show a detectable increase in their intensity after light excitation are (400), (040), and (220) (Fig. 2a). The (220) belongs to the {hh0} family and corresponds to the d-spacing of 3.2 Å (Pb-I bonds) along the octahedra diagonal (shown as the highlighted yellow rectangle in Fig. 2b), whereas the (400) and (040) Bragg peaks are in the {h00} family correlated with the d-spacing of 2.25 Å (half of octahedron length) along the edge of the octahedra (**a** and **b** directions, shown as cyan in Fig. 2b). On the contrary, higher orders of these two Bragg families such as (800), (330) and (550)), as well as other directions with reasonable signal ((310), (530), (750) and (1020), all exhibit a decrease in their intensities after photoexcitation. Details of these Bragg peak traces are illustrated in Fig. S3.  Furthermore, these transient intensity response scales monotonically with the pump fluence (1mJ/cm$^2$, estimated carrier density 1.3×10$^{13}$cm$^{-3}$), as shown in Fig. S4 for the representative Bragg peaks.

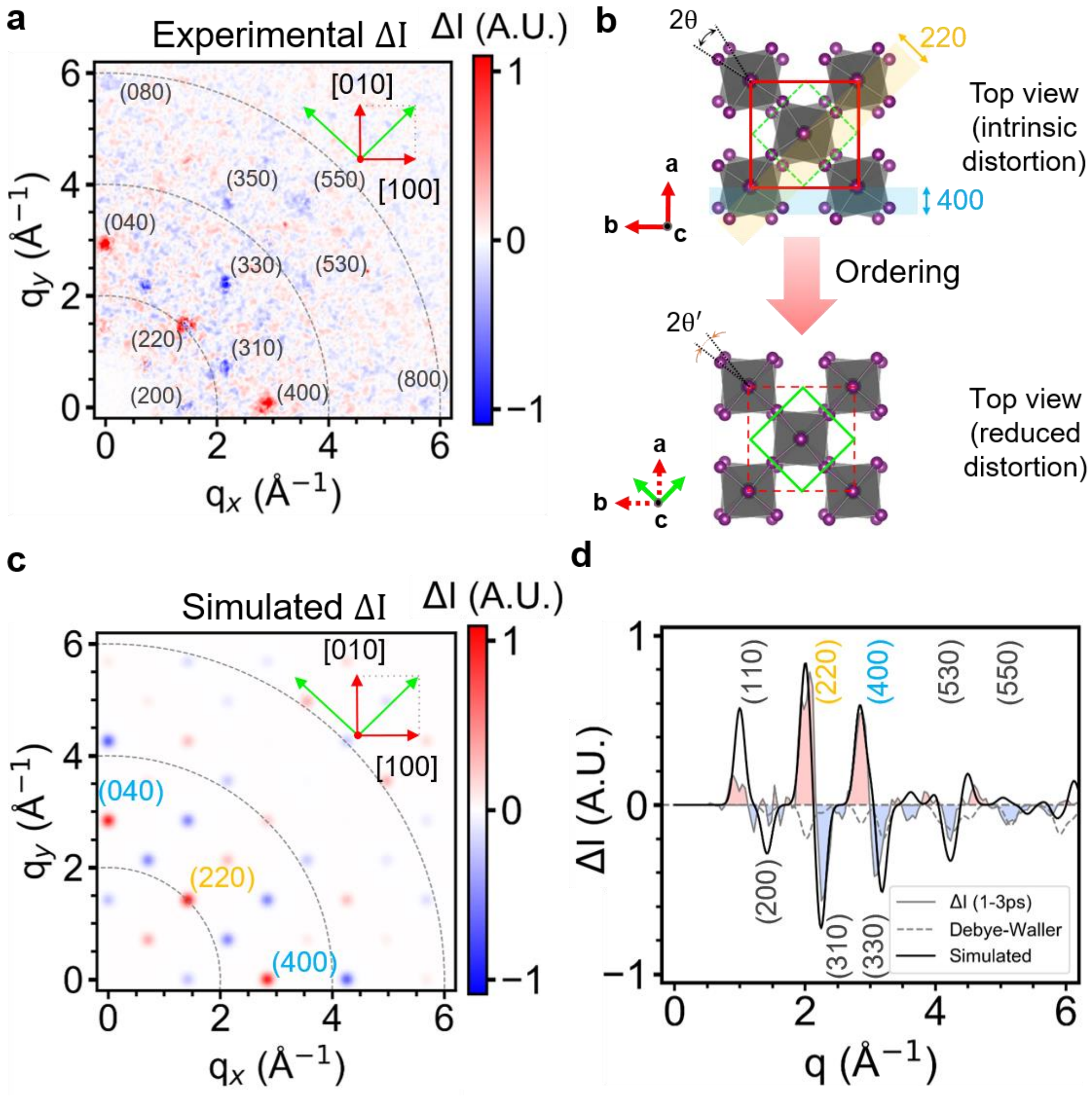

**Fig. 2 | Collective lattice ordering observed from UED response of DJ n=2 crystal. a,** Differential diffraction map at t=2 (±1) ps, acquired by subtracting the averaged response of 1~3 ps after light excitation by the response at rest (t≤0 ps). The intensity change ΔI (increase and decrease) of each specific Bragg peak (hk0) is indicated by red and blue colors, respectively. The map is zoomed in to the first quadrant ($0<q_x$, $q_y<6$Å$^{-1}$). **b,** Crystal structure at rest showing intrinsic antiferro-distortion (top) and simulated structure reducing the distortion (bottom), with disorder parameter θ defined as half of the interlayer octahedral rotations. For clarity the organic cations are omitted. **c,** Simulated differential diffraction map between the two crystal structures in Fig. 2b, showing intensity change (ΔI) of each Bragg peak after reducing the disorder parameter θ. **d,** Angular-integrated differential diffraction (filtering out diffuse scattering regions), comparing the experimental data (t=2 (±1) ps) (gray solid line and shadow), the expected Debye-Waller response (gray dashed line), and simulated signal acquired from Fig. 2c (black solid line). Note that the {110} peaks are partially filtered by the direct beam blocker (Fig. 1c).

We note that opposite to our results, classical semiconductors (such as Si and GaAs), including inorganic 2D materials (for example, transition metal dichalcogenides such as $MoS_2$ and $MoSe_2$), exhibit in most cases a decrease in Bragg peak intensities after photoexcitation.[22,27,28] In layered 2D materials an increase of Bragg peaks has been reported, which results from a suppression of charge density waves leading to a change of crystal symmetry,[29] however no evidence of charge density waves has been reported in halide perovskites. Additionally, large multiple scattering effects in monocrystalline samples can also spuriously induce unexpected increase of Bragg peak intensities upon photoexcitation.[30] However, our estimation of extinction distance of DJ n=2 perovskite is much larger than its sample thickness, implying that the multiple scattering effects are not likely to occur for our 2D perovskites, especially under MeV electron beams, whose cross sections are much lower than KeV electrons (see Supplementary discussion 3). In classical materials, the UED signal is attributed to a typical Debye-Waller response, which corresponds to an energy transfer from hot carriers to the low-frequency vibrational density of states, which scale as $|q|^2$.[22,31] Recent study on hybrid three-dimensional (3D) perovskites ($MAPbI_3$) also suggests a Debye-Waller like response, where octahedral rotational disorder induces additional structural deformations.[23] The counterintuitive and opposite behavior of light-induced ordering in some specific crystal directions in DJ n=2 crystals implies that there must exist another competing mechanism, via which the photogenerated hot carriers strongly interact with the lattice resulting in transient structural change, instead of thermal activation of the phonon bath leading to lattice heating. The increase in the intensities of specific Bragg peaks, thereby counteracting the Debye-Waller effect at short time scale and enhancing the lattice ordering in specific directions.

## Mechanism of transient lattice ordering

Qualitative analysis of the diffraction peaks in Fig. 2a suggests that the light-induced transformation may be related to the in-plane lattice ordering along specific directions. Therefore, we hypothesize that this observed anisotropic lattice ordering originates from the reduction of antiferro-distortions, a specific type of structural deformation involving the rotations of octahedra, which are widely expected in perovskites materials.[9,23,32] To quantify this distortion, we define the in-plane distortion angle θ as the half of the interlayer Pb-I bonds tilt (Fig. 2b), where 2θ is around 24° for the intrinsic structural distortion.[24] To further testify this hypothesis of octahedral rotations, we slightly reduced the distortion parameter θ to match the extent of Bragg peak increase in Fig. 2a, and simulated the differential diffraction pattern with respect to the crystal structure in the dark ($\theta_0$ = 12°). Compared with the relative intensity changes at short time, the optimal angle tilt is found to be ~0.25° (details of simulation shown in Supplementary discussion 4 and Fig. S5), and the simulated differential diffraction map in Fig. 2c. Compared to the distorted crystal structure at rest, the new structure reduces the adjacent octahedral tilt along the stacking axis (antiferro-distortion), and better aligns the iodine atoms for both octahedral edge {h00} and diagonal directions {hh0}, especially the distance spacings of Pb-I bonds {220} and half of octahedral edge lengths {400} (yellow and cyan in Fig. 2b). This structural change intensifies the in-plane Bragg peaks including {400} and {200}, consistent with the experimental observation at short time (Fig. 2a). Furthermore, we note that the expected intensity increase is not observed for the higher orders of the Bragg peaks in these two families, such as {800}, {330} and {550}, (Fig. 2a). We attribute this deviation to the combination of both the light-induced ordering and the Debye-Waller response. However, because the Debye-Waller effect scales with a $|q|^2$ dependence, it dominates and thus neutralizes the effect of ordering for the higher order Bragg peaks. We note that the response of {110} peaks are partially blocked by the direct electron beam filter, resulting a lower-than expected intensity increase in Fig. 2d. To quantitively compare these intensity changes, we also perform an angular integration at constant |q| of these differential diffraction images (t=2ps and the simulated one after ordering) (Fig. 2d). The experimental and simulated results are in fair agreement, especially in the low q-value region ($1 \text{ Å}^{-1} < |q| < 3 \text{ Å}^{-1}$) where the Debye-Waller effect is weaker. Therefore, these observations support the hypothesis that the light-induced ordering in DJ n=2, is due to the collective lattice reorganization that reduces the octahedral rotations (Fig. 2b). As the in-plane octahedra antiferro-distortion of the perovskite lattice at rest is non-polar and the light-induced atomic motions are essentially affecting the magnitude of the distortion (Fig. 2a), our results imply that carrier coupling to the related optical phonon mode is associated with a non-polar deformation potential mechanism.[6] This analysis is further confirmed by our calculations of vibrational modes, which reveals that antiferro-distortive phonon modes associated with the octahedra rotations are predicted for the crystal structure at rest (Fig. S6).

Next, in order to gain a deeper insight into the time evolution of the light-induced structural dynamics in 2D hybrid perovskites, we performed a full analysis (up to 100 ps) of the time-dependent intensity changes of the diffraction images after optical excitation. The time traces of the two Bragg peaks {400} and {530} are displayed in Fig. 3a, where the {400} peaks exhibit a significant faster rise-time (~1ps, Fig. S3), revealing a diverse response in Bragg peak dynamics besides the intensity change. Furthermore, an overall increase of diffuse scatterings in-between the Bragg peaks spots is visualized, in the differential diffraction images as shown in Fig. 3b. Compared between short-time (t=2ps) and long-time delays (t=65ps), the regions in-between the Bragg peaks spots appear redder as most of the Bragg peaks intensifies toward a negative response. The evolution of the integrated diffuse scattering intensities yields an average increase with ~0.5% (Fig. S3), suggesting a broad energy transfer process to the phonon bath, along with the Debye Waller effect specified at Bragg peak locations.

To further quantify the different types of lattice dynamics, we fit each Bragg peak intensity curve with a single exponential function, yielding a rise-time constant $\tau$ and relative intensity change $\Delta = (I - I_0)/I_0$, where $I_0$ and I correspond to the diffraction intensity at rest and after delay time t, respectively. Here, the parameter $\Delta$ represents the total peak intensity change due to carrier-lattice interactions and $\tau$ the time to achieve 1/e intensity change as shown in Fig. S7a and S7b, respectively. The maximum positive change (increase in intensity) is 4% and the negative change (decrease in intensity) is 12% (detailed plots in Fig. S3a). The time constant map shows a clear disparity between the dynamics from different Bragg peak groups. First, the Bragg peaks with positively changing intensity for which the light-induced lattice ordering mechanism dominates ({400} and {040}), exhibit a very fast time constant of 1 ps or shorter, whereas the peaks with negative intensity response yield a longer time constant of the order of 5 ps. Similar time constants have also been reported in 3D perovskites and have been attributed to the Debye-Waller like response originating from the rotational disorder of the iodine atoms in the perovskite octahedra, leading exclusively to a decrease in peak intensities.[23] A modification of our model including the Debye-Waller effect at long time scale is shown to give a better prediction of intensity changes at large q range than the lattice ordering effect (Fig. S8), confirming the existence of multiple energy transfer process in 2D perovskites. Second, both the intensity changes and time constants show a significant anisotropy with respect to directions in the DJ n=2 crystals, which reflects the lack of a tetragonal axis in the monoclinic centered structure (space group Cc n°9) of the crystals at rest because of the geometry and structure of the 4AMP cation. We also note that the intensity traces of some Bragg peaks such as {330} and {550} show a long recovery time (>30 ps, Fig. S3a), which could be related to other heat transfer mechanisms and phonon-phonon interactions.[27]

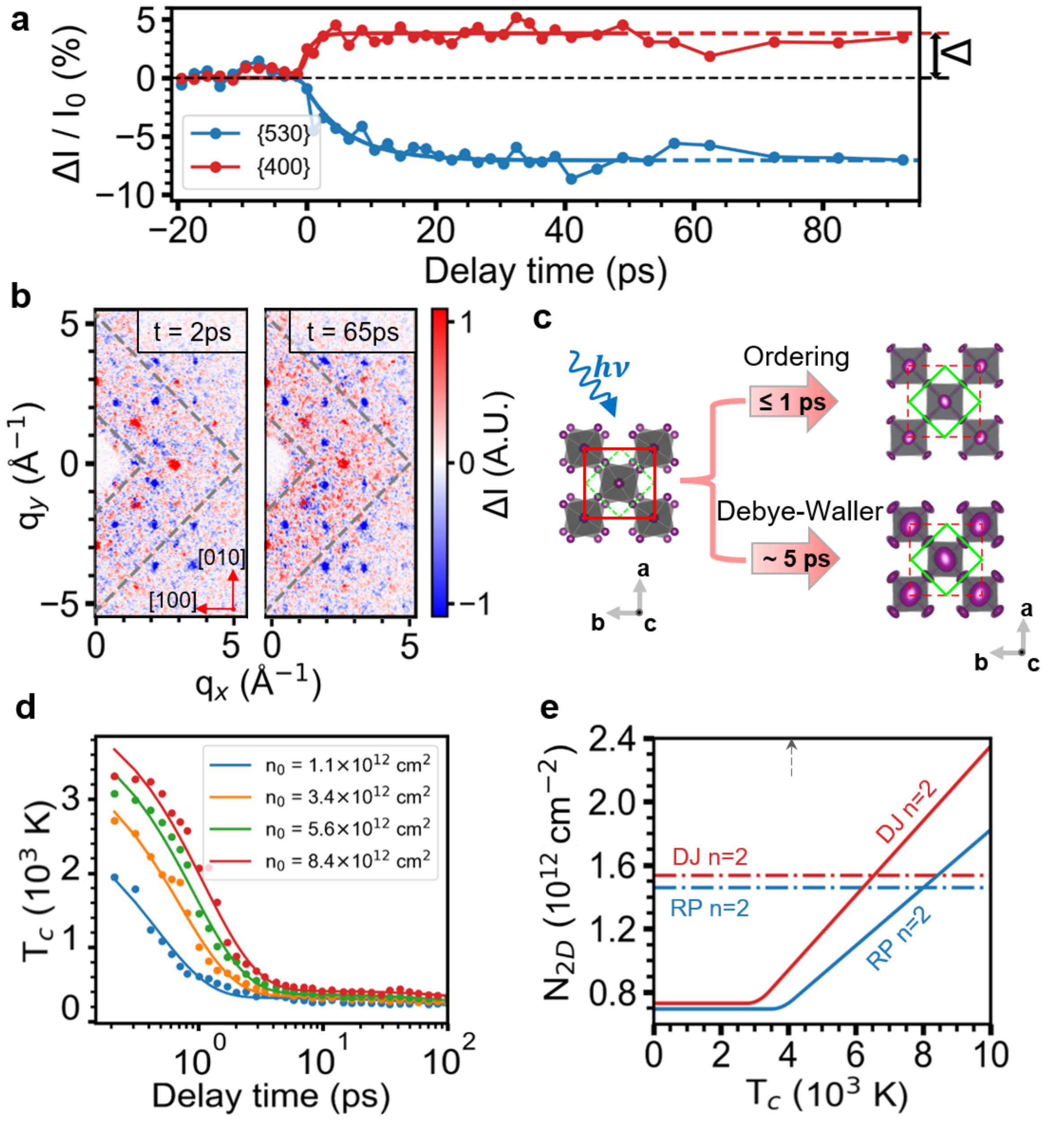

**Fig. 3 | Lattice (in-plane) and carrier response of DJ n=2 crystal under light excitation.** **a,** Intensity dynamic plots of peak families {400} and {530}. **b,** 2D differential diffraction map averaged at t = 2ps (left half) and t = 65ps (right half), with blue for intensity decrease and red for increase. The dashed region indicates the area with the most increase of diffuse scatterings. **c,** Schematics of light-induced reduction of antiferro-distortion and thermally activated atomic displacement from Debye-Waller effect. **d**, Evolution of carrier temperatures ($T_c$) as a function of delay times at different excitation densities ($1.1\sim8.4\times10^{12}$ $cm^{-2}$). **e**, Estimation of saturation densities ($N_{2D}$) for DJ n=2 (red) and RP n=2 (blue) crystals at different temperatures. The saturation densities due to electron-hole plasma and exciton gas are shown in solid and dashed lines, respectively.

All these observations from UED suggest that there are multiple competing mechanisms at play, which attest to the different origins and time scales of carrier-lattice interactions, as illustrated in Fig. 3d. We propose that the light-induced lattice ordering, which was attributed to the reduction of the in-plane lattice antiferro-distortion at ultrafast time scale (≤ 1 ps), and at longer timescales (~5 ps) a Debye-Waller like response. Here, the transient increase in the collective ordering in the DJ n=2 lattice may be viewed as an incomplete transformation towards a completely ordered phase. Indeed, differential calorimetry measurements on DJ n=2 single crystals do not show any evidence of a temperature-induced phase transition above ambient temperature toward a perfectly ordered phase (Fig. S9). Besides, temperature dependent XRD exhibits no sign of phase transition, and rather results in a global decrease of the Bragg peak intensities due to a Debye-Waller effect and linear thermal expansion (Fig. S10). Additionally, the perfectly undistorted 2D perovskite reference phase has only been observed in the fully inorganic Cs-based $Cs_2PbI_2Cl_2$ 2D perovskite at room temperature,[33] which emphasize the importance of the anisotropy of the 4AMP organic cations in the structure for the lattice distortion in hybrid 2D perovskites. On the contrary, undistorted phases are common for hybrid 3D perovskite compounds at room temperature and therefore any ultrafast light-induced ordering is lacking.[34]

## Ultrafast carrier-lattice interactions triggered by dense electron-hole plasma

In order to gain a complete picture of energy transfer and understand the structural dynamics observed in UED measurements, we performed transient absorption (TA) measurements in the same temporal window with the similar excitation fluences as the UED experiments on DJ n=2 crystal, (SI discussion 5 and Methods). By monitoring the transmission variations following above-band optical pump et 3.1eV, we estimated the evolution of effective temperature ($T_c$) of hot-carrier gas, by fitting the high-energy tail near the band edge with the Maxwell-Boltzmann distribution. (analysis details in supplementary discussion 5 and Fig. S11).[35] The hot-carrier dynamics for various fluences are displayed in Fig. 3c. We observe a fast build-up time of initial carrier temperature within hundreds of femtoseconds, corresponding to carrier-carrier scattering before quasi-equilibrium distribution of hot-carriers. The carrier cooling time is estimated to be sub-ps to 1ps, which increases monotonically with excitation fluence. This carrier cooling time constant is consistent with previous reports on DJ perovskites poly-crystalline films ((4AMP) $PbI_4$).[36] The extracted hot-carrier cooling time was 1.1 ps at $8.4\times10^{12}$cm$^{-2}$, which also corresponds to the rise time of (400) Bragg peaks that exhibit lattice ordering, thus hinting at correlated processes that govern the structural change and the carrier dynamics measured using ultrafast spectroscopy.

To elucidate the microscopic origin of ultrafast lattice ordering, and how the role of photoexcited carriers in activating the reduction of antiferro-distortion, we performed an in-depth analysis of the excitation regime (electron-hole pair plasma vs exciton gas) of DJ n=2 samples in UED. In order to achieve this, we followed the seminal analysis proposed by Schmitt-Rink, Chemla and Miller for excitonic non-linear optical effects in 2D semiconductors.[19,20] First, at a fluence of 1 mJ/cm$^2$ in UED experiment, a high electron-hole pair density of $1.3\times10^{13}$ cm$^2$ is generated. The effective temperature of initial carrier gas is estimated to be 4100K, extrapolated from the carrier temperature plotted in Fig. 3c. Next, we computed the saturation densities for excitonic responses in DJ n=2 perovskites, by taking into account the exciton binding energy and Bohr radius of 1s exciton resonance, and comparing these values with RP n=2 compound (see supplementary discussion 6).[19,37] A summary of these saturation densities at different carrier temperatures is plotted in Fig. 3e. Here, the dashed horizontal lines correspond to the saturation due to the exciton gas (Mott transition), above which photoexcited carriers are not described by excitons but by a hot electron-hole plasma.[19] The solid line draws the saturation by electron-hole plasma at low and high temperature ranges. As analyzed by Schmitt-Rink, Chemla and Miller, a cold exciton gas is more efficient in saturating the exciton resonance than a hot carrier plasma. The very high plasma pair density of $1.3\times10^{13}$cm$^{-2}$ (for 1 mJ/cm$^2$ in the UED experiment) nonetheless exceeds the saturation densities for both hot carrier plasma and cold exciton gas regimes by one order of magnitude. Therefore, we conclude that the observed energy transfer at short time in the UED experiment is rather related to the coupling of a hot and dense carrier plasma to the lattice, instead of exciton polaronic couplings as reported in other studies at low temperature and low excitation power.[18]

In the case of a dense electron-hole plasma, strong transient lattice distortions different from Debye-Waller effects can be induced in semiconductors. For example, transient strain pulses can be generated by supersonically propagating plasmas, traveling deep into crystal bulk beyond the speed of sound.[38,39] Moreover, complex lattice organizations have been also recently observed, such as the suppression of the ferroelectric instability $KaTO_3$,[40] and the activation of the antiferro-distorsive rotation of octahedra in $SrTiO_3$.[41] The last case, which was interpreted on the basis of a photodoping-induced modification of the structural soft-mode potential, might be closely related to our present study on 2D perovskites. Furthermore, it was shown that the ultrafast dynamics (~0.2 ps) of the observed antiferro-distortion in $SrTiO_3$ is non-thermally driven by changes of the phonon potential.[41] These results and analysis have enabled us to propose a complete picture involving carrier-lattice interactions in short time, where high-density injection of carriers generates a hot and dense electron-hole plasma. The electron-hole plasma that produces a strong modification of the lattice dielectric properties on one hand results in the screening of bound electron-hole (exciton) pair formation, and on the other hand modifies the interatomic force constants. This last mechanism of plasma-induced modulation of the interatomic force must be considered because halide perovskites are strongly ionic materials like $SrTiO_3$. The resulting modification of the crystal cohesive energy, manifests as a strong lattice reorganization with antiferro-distortive rotation of octahedra in 2D DJ perovskites. In parallel, a more classical decay of the plasma energy occurs through electron-phonon interactions and observed in structural dynamics associated with a Debye-Waller effect. This classical process is mainly associated to an initial activation of optical phonons, that decay later toward acoustic phonons producing coherent acoustic oscillations (as shown in Fig. S11e).[36]

## Role of organic cations and structural phases of 2D perovskites

Finally, in order to understand if the mechanism of light-induced transformation is generally applicable to the other 2D perovskites with different organic cations and structural phases, we perform UED experiments on three other hybrid 2D perovskites: 4AMP-$MA_2Pb_3I_{10}$ (DJ n=3) exhibiting the same composition as DJ n=2 but a thicker perovskite layer, $BA_2MAPb_2I_7$ (RP n=2) of Ruddlesden-Popper (RP) phase type with same perovskite layer as DJ n=2 but different spacer organic cations (with BA butylammonium), and $BA_2MA_3Pb_4I_{13}$ (RP n=4) with a different perovskite layer and spacer organic cations. The structures of these 2D perovskites are illustrated in Fig. 4a and Fig. 4b, respectively. The experimental UED static diffraction pattern of DJ n=3 (shown in Fig. S12) resembles that of DJ n=2, suggesting that changing the perovskite layer thickness for the DJ 2D perovskites has a negligible effect on the in-plane diffraction pattern.[24] From the response of DJ n=3 (shown in Fig. S13), we also observe similar intensity increase for peak (040) and (004), as was observed for DJ n=2 (Fig. 2a). Furthermore, the rise time of these Bragg peaks are at similar time scale of ~1ps (Fig. S14). These analysis of the DJ n=2 and n=3 indicate that the phenomena of light-induced ordering at short time scales is generally applicable to the DJ type 2D perovskites. In contrast, the RP crystal no signature of fast rising peaks with positive intensity was observed (Fig. S14 to Fig. S16). The non-existence of light-induced ordering in RP phase could be attributed to the fact that the crystal structure is already highly ordered in RP perovskites for higher n-values (Fig. 4a and Fig. 4b).[6]

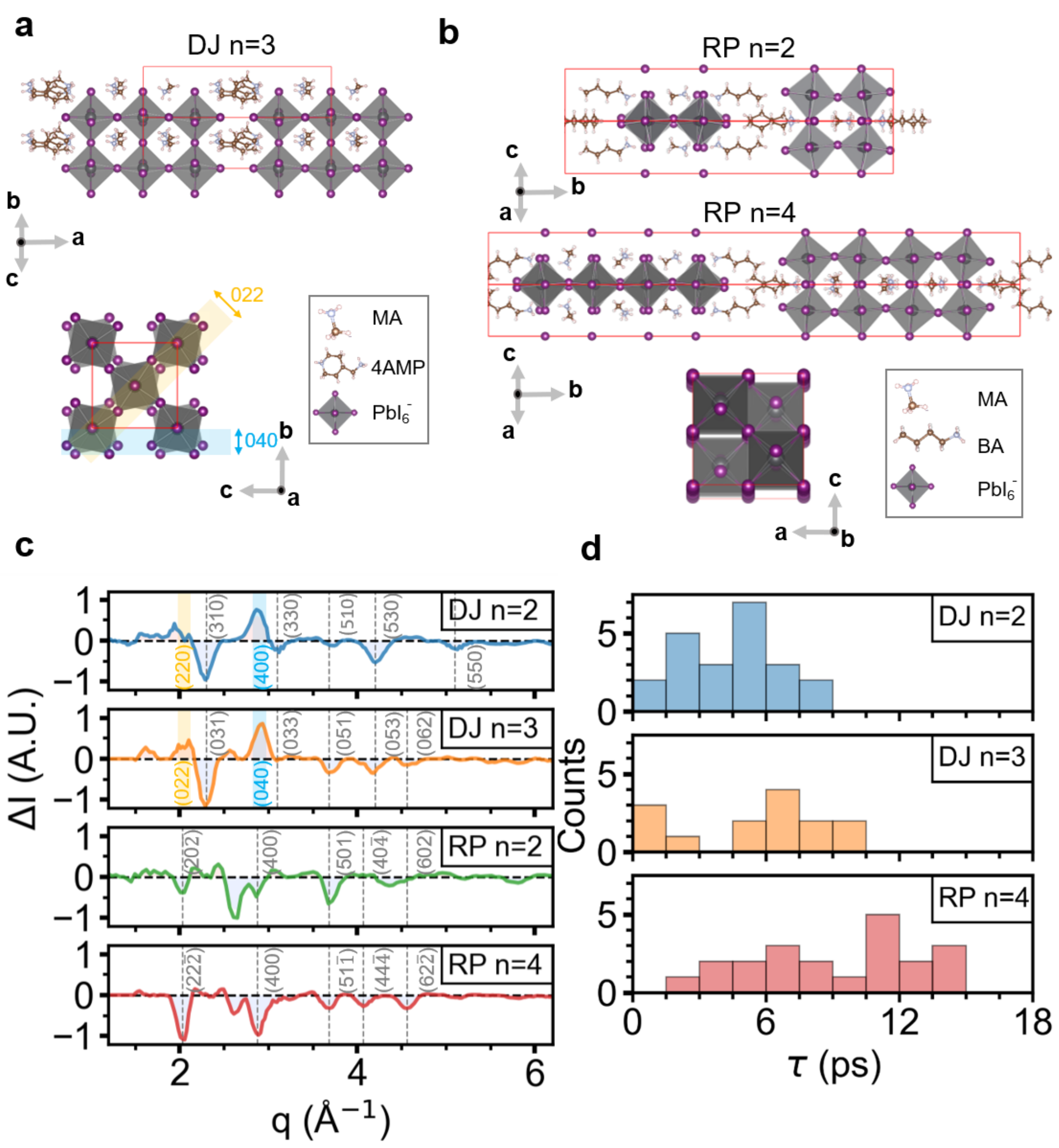

**Fig. 4 | Comparison between different 2D perovskite structures with different organic cations.** Crystal structures of the **a,** DJ n=3 and **b,** RP n=2 and n=4 perovskites. **c,** Angular integrated differential diffraction plot (20~40ps) for the DJ n=2, DJ n=3, RP n=2 and RP n=4 perovskites. The yellow and cyan colors label the Bragg peaks showing positive responses in DJ systems. **d,** Histogram of rise-time constants of DJ n=2, DJ n=3 and RP n=4 structures.

In order to quantify the different behaviors between the DJ and the RP phase 2D perovskites we evaluated the differential intensity and the time constant of the relevant Bragg peaks. First, we analyzed the differential intensity comparison between DJ n=2 and DJ n=3 as illustrated in Fig. 4c. Both DJ n=2 and n=3 exhibit similar behaviors. We observed an increase in the intensity of the in-plane Bragg peaks for DJ n=2 {400} and DJ n=3 {040}, and a decrease for the intermediate peaks {310}, {330}, {510} and {530}, consistent with the Debye-Waller effect. However, in sharp contrast to the DJ case, all the observed Bragg peaks in RP n=2 and n=4 yield a global decrease in intensity. Furthermore, comparison of the histograms of the time constant $\tau$ for DJ n=2, DJ n=3, and RP n=4 (Fig. 4d, see also Fig. S17 for the corresponding scatter plots), shows a generally slower response (up to 15 ps) of RP perovskites as compared to the DJ crystals. As similarly observed in DJ n=2, an increased diffuse scattering at longer time scales (Fig. S13, Fig. S15 and Fig. S16) indicates a thermal energy transfer to phonon populations, with similar time constant as the Debye-Waller decrease.

These results demonstrate that there is a fundamental difference in the mechanism of carrier-lattice coupling and structural dynamics between the DJ and RP 2D perovskite phases, which implies that the competing mechanism of ultrafast lattice ordering during an antiferro-distortive phase transformation is mitigated in the RP phase. As described in Fig. 4b, the diffraction patterns of RP n=2 and n=4 at rest are consistent with the static crystal structures reported previously, which exhibits almost no antiferro-distortions in plane,[42] contrary to DJ n=2 and n=3 (Fig. 1d and 4a)[24]. Therefore, in the RP case the initial ordering step (Fig. 3d) is by-passed and the energy of the hot carriers is directly released to the lattice without the initial resonant energy transfer. We also note that the RP n=4 case is more consistent with the results on 3D $MAPbI_3$, which only shows the Debye-Waller like effect with similar time scales (~ 10 ps).[23] In addition, the slow response time in RP systems arises from the differences in the lattice softness between the BA and the DJ 2D perovskites. The BA molecules reduce the rigidity of the 2D lattice by comparison to the DJ case, leading to an enhancement of the lower-energy part of the vibrational density of states.[43] It was indeed shown experimentally to enhance the deformation potential mechanism in BA compounds, leading to carrier trapping.[5] This results is also consistent with a recent high-resolution resonant impulsive stimulated Raman spectroscopy,[12] showing that the vibrational relaxation in monolayered bromide perovskites formed from flexible alkyl-amines as organic barriers is fast and relatively independent of the lattice temperature, while a resonant effect is observed for aromatic amines. These results comparing the nature of carrier lattice interaction in 2D perovskites with different organic spacer layers provides clear design principle in controlling of carrier dynamics in 2D perovskites-based devices.

Thus, in summary, our work elucidates that the photoexcited carrier relaxation depends on a competition between: i) a fast lattice ordering induced by hot electron-hole plasma, which strongly modulates the crystal cohesive energy by reducing the lattice antiferro-distortions; and ii) more classical ones related to a decay toward the low-energy part of the vibrational density of states. The strongly contrasting behaviors observed by changing the interlayer cations and thereby the structural phase of 2D perovskites demonstrates that UED measurements can be further extended systematically to the study of different perovskite compounds, such as organic cations and halides, providing an ideal platform of controlling structure-energy dissipation channels in organic-inorganic materials relevant for a wide range of novel applications including hot-carrier photovoltaics and photocatalysis and ultrafast optical communication devices.

## Methods

### Materials

The materials used for 2D perovskites (DJ and RP) precursors were purchased from Sigma Aldrich, including methylamine hydrochloride (MACl), lead oxide (PbO), 4-aminomethyl piperidine (4AMP), butylamine (BA), Hypophosphorous acid ($H_3PO_2$), and hydroiodic (HI). Methylamine iodide (MAI) was purchased from Greatcell Solar.

### Single crystal synthesis and transfer

*Single Crystal Synthesis*: BA n=2,4 perovskites solution was synthesized by adopting the previously reported procedure, using 0.4 times the scale.[24] 4AMP n=2,3 perovskites solution was synthesized by adopting the previously reported procedure, using 0.5 times the scale.[42] Glass was used as the substrate for the 2D perovskite growth. Glass substrates were cut into 2.5 cm × 2.5 cm squares, then cleaned in soapy water, isopropanol, acetone by ultrasonication for 20 min each, and then were dried using an argon gun. The substrates were transferred into a UV-Ozone cleaner, and cleaned for 20 min. 10 μL of the diluted solution was deposited dropwise onto the glass surface, another glass was put on top to fully cover the bottom glass and dried overnight at 80 °C for BA and 75 °C for 4AMP. Thin single crystal grew spontaneously as the solvent evaporated.

*Crystal Exfoliation and Transfer*: Large-flake single-crystal perovskites were mechanically exfoliated and then transferred onto 100 × 100μm TEM transmission window with silicon nitride membranes. All samples were prepared and sealed in Argon gas until UED measurements. A comparison of diffraction images between first and last scans suggests no new peaks showing, confirming no degradation during measurement period (Fig. S18).

**Sample characterization**

*Optical absorbance:* The optical absorbance of the single crystals was determined based on confocal microscopy, by focusing a broad-spectrum white light beam (Thorlabs SOLIS-3C) into a 50um spot and measuring the transmission spectrum through a TEM window. The spectrum was acquired using a spectrometer (Andor Kymera 328i) and CCD (Andor iDus 416).

*Differential scanning calorimetry (DSC):* Differential scanning calorimetry was performed from room temperature to 150℃ with a ramp rate of 15℃/min. The heat capacity of the sample was calculated from the heat flow (W/g) and was calibrated using a sapphire sample as a reference.

**Ultrafast electron diffraction (UED)**

UED experiments were performed at SLAC National Accelerator Laboratory. Electrons were accelerated to 3.7 MeV by Klystron with ~150fs temporal width. Normal incident pulses were diffracted by the sample in transmission geometry on TEM transmission window, and diffracted intensities were captured by EMCCD. The electron pules were synchronized with a laser beam (3.1eV, ~75fs) at the frequency of 180Hz, with temporal delay typically from -20ps to 100ps. The delay time was tuned by adjusting the relative path length difference using a delay stage. To reduce the confounding effect of mechanical shift and possible crystal structural change during time evolution (~hrs.), the sequence of stage locations was configured randomly during each scan. Each typical UED run contained 20~30 pump-probe scans and each diffraction image was averaged over ~ 8000 shots to reduce statistical fluctuations.

*UED Bragg peaks indexing:* The Bragg peaks of the measured samples (DJ n=2, DJ n=3, RP n=2 and RP n=4) were indexed by comparing the diffraction patterns with the simulated diffraction images calculated via SingleCrystal software (http://crystalmaker.com/singlecrystal/), using structure files (.cif) from Mao et al.[24] These comparisons were shown in Fig. S2 (DJ n=2), Fig. S12 (DJ n=3), Fig. S19 (RP n=2) and Fig. S20 (RP n=4).

*UED data processing:* A stack of diffraction maps at different time step was collected for each sample measurement. Each diffraction image was background-removed by CCD dark count and normalized by total diffraction intensity. To extract the time traces of relative intensity response of each Bragg peak ($\Delta I(t)/I_0$), we first integrated the intensity within ROI centered at each peak (0.2Å in diameter), and then normalized the intensity curve by averaging over negative delays (t<0). Time traces of each Bragg peaks were fitted globally by a single exponential curve:

$$\Delta I(t)/I_0 = \begin{cases} 0 & , \quad t < t_0 \\ \Delta\left(\exp\left(\frac{t-t_0}{\tau}\right) - 1\right), & t \geq t_0 \end{cases}$$

Where $\Delta$ stands for intensity change (%), $\tau$ for time constant (ps), and $t_0$ for time zero offset (calibrated by the response of bismuth crystal and fixed globally). Fitting results were plotted in Fig. S7 (DJ n=2) and Fig. S17 (DJ n=3 and RP n=4). Results of selected peak families of DJ n=2 are displayed in Fig. S3. Error intervals of the fitting parameters $\Delta$ and $\tau$ were estimated from one standard deviation calculated from the square root of the covariance matrix diagonal. Fitting methods were based on least squares optimization performed by 'curve_fitting' function using Python (version 3.7.4).

**Transient absorption spectroscopy (TA)**

Transient absorption (TA) experiments for 2D perovskites were done by using a 400 nm pump laser and a supercontinuum white light probe laser ranging from 450 to 750 nm. The pump laser was obtained from a 2$^{nd}$ harmonic generator (Ultrafast System) derived by a fundamental 60 fs-pulsed laser (Coherent, Inc.) at 800 nm with a repetition rate of 1 kHz. The 1 kHz pump laser passed through a chopper to yield a 500 Hz frequency. The pump fluence was controlled within 0.1~0.75 mJ/cm$^2$. The supercontinuum white light laser was generated by introducing the fundamental laser into a nonlinear crystal, CaF2 (Ultrafast System). Pump and probe laser pulses were temporally and spatially overlapped and focused on the 2D perovskites. 2D perovskites were performed in ambient air condition with DJ n=2 sample sealed in ~50nm thick PMMA.

**Phonon simulation methods**

*Phonon modes simulation:* these were based on periodic Density Functional Theory (DFT) calculations, performed within Linear Combination of Atomic Orbital (LCAO) formalism, as implemented in the CRYSTAL17 code.[44] Electronic ground state was computed within the GGA approximation, employing the PBE exchange-correlation functional,[45] along with double split basis set quality and core potentials for heavy Pb and I atoms. Dispersion D3 scheme was included to account for van-der Waals interactions,[46] as these are expected to play an important role in the packing of the organic component. As it is not currently implemented in CRYSTAL17, Spin-Orbit-Coupling (SOC) has been is neglected in the present calculations. Nevertheless, SOC is not expected to heavily affect the ion dynamics, with current computational set-up providing vibrational frequencies in nice agreement with experimental data, for the reference orthorhombic phase of the 3D methylammonium lead iodide perovskite.[47]

Calculations have been performed on the reference crystalline structure of the DJ n=2 system from Ref.[24], adopting 2x4x4 sampling of the first Brillouin zone, in the Monkhorst-Pack scheme[48] (with the less dense sampling associated to the longer plane-stacking direction). Crystalline structure was fully relaxed with only constraint related to the native space group symmetry (Cc n°9). The hessian was estimated at the Γ point of the Brillouin zone, via finite displacements,[49] resulting in zero negative frequencies. Resulting normal modes were used as special coordinates along which to distort the crystal structure and to evaluate the change in the electron diffraction pattern, as exemplified in Fig. S6.

*Electron diffraction simulation:* snapshots of the crystalline structure as distorted along all the computed normal modes were then provided as crystallographic information files (.cif) and used to simulate electron diffraction intensities using SingleCrystal software. The simulated 2D patterns and 1D azimuthally integrated plots were reconstructed by convolving the localized Bragg peaks with gaussian shape with FWHM of 0.17Å.

## Data availability

The authors declare that the main data supporting the findings of this study are available within the article and its Supplementary information. Extra data are available from the authors upon request.

**Acknowledgments:** The work at Rice University was supported by start-up funds under the molecular nanotechnology initiative and also the DOE-EERE 2022-1652 program. J.E. acknowledges the financial support from the Institut Universitaire de France. Work at Northwestern was supported by the Office of Naval Research (ONR) under grant N00014-20-1-2725. The experiment was performed at SLAC MeV-UED facility, which is supported in part by the DOE BES SUF Division Accelerator & Detector R&D program, the LCLS Facility, and SLAC under contract Nos. DE-AC02-05-CH11231 and DE-AC02-76SF00515.

**Author contributions:** J.-C.B. and A.D.M. conceived and designed the experiment. J.H. and S.S. synthesized the perovskite single crystals, and W.L. prepared samples with the help of J.H. and H.Z. H.Z. performed optical characterizations. W.L., H.Z., S.S. and A.F. performed the UED experiments with the help of. A.A, M.-F.L, A.B., X.Z. and XJ.W.. H.Z. performed data analysis under the help of J.E and I.M, with guidance from J.-C.B. Phonon modeling was carried out by. C.Q.. W.Y. performed TA measurements. J.-C. B. and A. D. M. wrote the manuscript with inputs from everyone. All authors read the manuscript and agree to its contents, and all data are reported in the main text and supplemental materials.

**Competing interests:** The authors declare no competing interests.

## *Supplementary Information for*

# Direct visualization of ultrafast lattice ordering triggered by an electron-hole plasma in 2D perovskites

Hao Zhang[1,2], Wenbin Li[1,2], Joseph Essman[1], Claudio Quarti[3,4], Isaac Metcalf[5], Wei-Yi Chiang[6], Siraj Sidhik[1,5], Jin Hou[5], Austin Fehr[1], Andrew Attar[7], Ming-Fu Lin[7], Alexander Britz[7], Xiaozhe Shen[7], Stephan Link[6], Xijie Wang[7], Uwe Bergmann[8,9], Mercouri G. Kanatzidis[10], Claudine Katan[3], Jacky Even[11], Jean-Christophe Blancon[1]* and Aditya D. Mohite[1]*

**[1]Department of Chemical and Biomolecular Engineering, Rice University, Houston, Texas 77005, USA.**

**[2]Applied Physics Program, Smalley-Curl Institute, Rice University, Houston, TX, 77005, USA.**

**[3]Univ Rennes, ENSCR, INSA Rennes, CNRS, ISCR (Institut des Sciences Chimiques de Rennes) - UMR 6226, F-35000 Rennes, France.**

**[4]Laboratory for Chemistry of Novel Materials, Department of Chemistry, University of Mons, Place du Parc 20, 7000 Mons, Belgium**

**[5]Department of Materials Science and NanoEngineering, Rice University, Houston, TX, 77005, USA.**

**[6]Department of Chemistry, Rice University, Houston, Texas 77005, USA.**

**[7]SLAC National Accelerator Laboratory, Menlo Park, CA 94025, USA.**

**[8]PULSE Institute, SLAC National Accelerator Laboratory, Stanford University, Stanford, CA, USA**

**[9]Department of Physics, University of Wisconsin–Madison, Madison, WI 53706 USA.**

**[10]Department of Chemistry, Northwestern University, Evanston, Illinois 60208, USA.**

**[11]Univ Rennes, CNRS, Institut FOTON (Fonctions Optiques pour les Technologies de l'Information), UMR 6082, CNRS, INSA de Rennes, 35708 Rennes, France.**

***Correspondence: blanconjc@gmail.com and adm4@rice.edu**

## This PDF file includes:

Supplementary discussions 1-8

Figs. S1 to S21

Table S1

References

## Supplementary Discussions 1-8

**Supplementary discussion 1.** Crystal orientation of DJ n=2, DJ n=3, RP n=2 and RP n=4 crystals

The crystal orientation of each sample measured in UED was acquired by comparing the diffraction patterns with the simulated diffractions viewing at different directions.

For the DJ systems, the experimental patterns in Fig. S2b (DJ n=2) and Fig. S12b (DJ n=3) show great consistency with the simulated ones (Fig. S2a & Fig. S12a) viewed at stacking axis direction ([001] for DJ n=2, and [100] for DJ n=3), confirming the horizontal orientation of the DJ crystals, where the perovskite crystal layers are parallel to the substrate. The azimuthally integrated plots (Fig. S2c for DJ n=2, Fig. S12c for DJ n=3) also show good agreement in peak locations and convoluted intensities. The measured in-plane Bragg peaks can be properly indexed accordingly, including (hk0) for DJ n=2, and (0hk) for DJ n=3. The labeled peaks are shown in 2D patterns and azimuthally integrated plot.

For RP n=2 crystal, the simulated pattern along stacking axis [010] is shown in Fig. S19a, where all the peak locations agree with the experimental ones in Fig. S19b, confirming the major orientation of the RP n=2 crystal is horizontal, same as DJ crystals discussed above. Additionally, concentric rings are observed between q=2 $Å^{-1}$ and 3 $Å^{-1}$, as observed in Fig. S19b, suggesting the existence of poly-crystalline like crystal structure with mixed orientations. The azimuthally integrated plots (Fig. S19c) show good agreement in peak locations and convoluted intensities, with additional unassigned peaks at q=2.6 $Å^{-1}$ and 3.7 $Å^{-1}$. We've excluded the possibility of $PbI_2$ diffraction by comparing with the electron diffraction of $PbI_2$ thin film.[1] We attribute them to the possible perovskite diffraction from out-of-plane or mixed directions, or an unidentified phase of 2D perovskites, and have excluded them from the discussion of in-plane Bragg peaks responses of RP n=2 crystals.

For the RP n=4 crystal, the simulated pattern along stacking axis [010] is shown in Fig. S20e, with corresponding 1D azimuthally integrated plot in Fig. S20f. However, this pattern doesn't match with the experimental one (Fig. S20a) based on the following observations: 1) more peaks are observed in static UED patterns, such as q=2.2$Å^{-1}$ and 3.8 $Å^{-1}$ as shown in Fig. S20f; 2) anisotropic intensity distribution along $q_x$ and $q_y$ directions, which is different with Fig. S20e with $4^{th}$ fold symmetry. Therefore, we conclude the beam direction is not perfectly along the stacking axis as observed in DJ and RP n=2 crystals. As shown in Fig. S20c & Fig. S20d, we find that the [011] orientation gives the best match for RP n=4, with 7.8 degrees deviation from stacking axis [010], as sketched in Fig. S20b. As a result, some out-of-plane Bragg peaks can be detected as well, which may explain the anisotropic UED response between $q_x$ and $q_y$ directions, including the intensity change shown in Fig. S14 and the time constants shown in Fig. S15.

**Supplementary discussion 2.** Sample Thickness and Carrier Density Estimation

*Crystal thickness determination:* The thickness of the crystal was determined by comparing the absorbance spectrum with the estimated one calculated from absorption coefficient of 2D perovskites. Here we have referenced the absorption coefficient results obtained from ellipsometry by Song et al.[2] Applying Beer-Lambert's law and neglecting light scattering, the relation of crystal thickness $d$ and crystal absorbance is given by:

$$\frac{I}{I_0} = 10^{-OD} = e^{-\alpha\, d} \tag{S2.1}$$

Where OD is the absorbance in optical density determined by transmission ($I/I_0$), and $\alpha$ is absorption coefficient. The comparison of experiment and simulated absorbance spectrum is depicted in Fig. S1.

*Carrier density:* The light-injected carrier density in quantum well is estimated by:

$$n_0 = \frac{\Phi/(h\nu)\, A}{d/l_w} \tag{S2.2}$$

Where $\Phi$ stands for energy fluence of each pulse (mJ/cm$^2$), $h\nu$ stands for excitation phonon energy (3.1 eV), $A$ stands for absorbance of the sample, $d$ is the sample thickness, and $l_w$ is the width of quantum-well in 2D perovskites. Here, we've neglected the sample reflection and estimated the sample absorbance by transmission data in Fig. S1. As a result, the list of thickness and corresponding carrier densities of each measurement is shown in Table S1.

**Supplementary discussion 3.** Possible reasons of lattice ordering in DJ perovskites.

A. *Phase transition from thermal lattice heating.* One of the possible mechanisms that induces the collective lattice ordering in DJ perovskites is the thermally induced phase transition from local lattice heating. To investigate the thermal response of DJ n=2 crystals, we first estimate the temperature jump induced by the laser pump, then using the differential scanning calorimetry (DSC) and temperature dependent XRD results to exclude the possible phase transition effect.

a. *Estimation of temperature jump.*

The local lattice temperature increase is estimated by total input pump energy and samples' specific heat. For DJ n=2 crystals, a carrier density of $2.5\times10^{13}$ cm$^{-2}$ corresponds to a bulk carrier concentration of n=$1.4\times10^{20}$ cm$^{-3}$, which induces total energy into the lattice:

$$\Delta E = n \times (E_{pump} - E_g) = 1.4 \times 10^{20} \times (3.1 - 2.2)\ eVcm^{-3} = 20.16\ J/cm^3 \tag{S3.1}$$

The mole of the DJ n=2 crystal is calculated from crystal structure (4 molecules per unit cell):

$$m = \frac{1}{V_{UC}} \times \frac{4}{N_A} = 2.53 \times 10^{-3} mol/cm^3 \tag{S3.2}$$

The specific heat of DJ n=2 at ~300K is estimated from DSC measurement, as shown in Fig. S9:

$$C_p = 0.54\ J\ K^{-1}g^{-1} = 728\ J\ K^{-1}mol^{-1} \tag{S3.3}$$

And the corresponding estimated temperature jump is:

$$\Delta T = \frac{\Delta E}{m * C_p} \sim 11K \tag{S3.4}$$

b. *Differential scanning calorimetry (DSC) results*.

To investigate possible thermal phase transition induced by lattice heating, differential scanning calorimetry (DSC) was performed from room temperature to 150℃, as suggested in Fig. S9. From the DSC curve, no significant phase transition is observed. Based on over estimation of temperature increase induced by light (SI discussion 5), the local temperature increase in the crystal lattice is below the 15K and no possible phase transition is observed or reported in this range.3 Therefore, we conclude the collective lattice ordering in light-induced in ultrafast scale, instead of thermally induced caused by lattice heating.

c. *Temperature dependent XRD.*

Additionally, we performed temperature XRD from room temperature to 55°C, data shown in Fig. S10. As we can see, most of the Bragg peak intensities go through a slight decrease with increasing temperature due to the presence of Debye Waller effect.

Here we focus on the in-plane Bragg peaks, primarily related to the edges and diagonal of the perovskite octahedral, which is our main focus in-plane diffraction patterns in DJ n=2. Fig. S10 (b)(c)(d) plot the peak profile variation with respect to the sample at room temperature, zoomed at q- regions near (220) and (400) respectively. As we can see, both (220) and (400) peaks exhibit slight intensity decrease, which is contradictory with the UED response.

Besides the intensity variations from observed from temperature XRD, we also observed a global peak shifting (upto 0.15%, Fig. S10 (f)). A global shifting of Bragg peaks toward low-q values is generally expected from the thermal expansion. However, this behavior of Bragg peaks shifting is not resolved in the UED. We attribute the absence of thermal expansion signature in UED diffraction patterns to the low q resolution of the UED measurements (~0.17$Å^{-1}$).

B. *Multiple scattering effect.* Another possible mechanism would be the possible multiple scattering effect in single-crystalline samples, as reported by Vallejo et al using keV electron beams.[4]. In fact, as pointed out by Vallejo et al, the possibility of having multiple scattering will rely on a characteristic length known as the extinction distance ($\xi_g$) of the samples, which is strongly affected by the electron beam energy, crystal structure and elements of material (Williams & Carter 2008[5]) as described by the following formula,

$$\xi_{\boldsymbol{g}} = \frac{\pi V cos\theta_B}{\lambda F_{\boldsymbol{g}}} \tag{S3.5}$$

Where $V$ is the volume of the unit cell, $\theta_B$ is Bragg angle, $\lambda$ is electron wavelength, and $F_g$ is the structure factor at diffraction vector $\boldsymbol{g}$.

The calculation of $V$, $\lambda$, and $\theta_B$ is straightforward. For high energy electrons $\theta_B$ is fairly close to 0, since the electron wavelength $\lambda$ is much smaller than lattice spacing ($\theta_B$=0.5 deg for Si (220) at 100keV) and therefore $\cos\theta_B$~1. The $F_g$ is estimated by the summation of all the atomic scattering factors with different atomic locations within the unit cell (Williams & Carter 2008 [5], Page 223):

$$F_g = F_{hkl} = \sum_{i=1}^{n} f_i^{(e)}\, e^{-2\pi i(hx_i + ky_i + lz_i)} \tag{S3.6}$$

The atomic scattering factor is a function of scattering vector s ($s = \sin\theta\,/\lambda$). and a common approximation is fitted with the following formula[6]:

$$f^{(e)}(s) = \sum_j a_j \exp(-b_j s^2) \tag{S3.7}$$

The coefficients of $a_j$, $b_j$ for each atoms are listed in Table 1 in Peng et al[6]. For ionized atoms (such as Pb2+ and I- ), additional term is included[7].

Based on the information above, we have estimated the extinction distance for Si under 100keV and DJ n=2 perovskites under 3.7MeV, which are given by:

$$\xi_{\boldsymbol{Si}|(\boldsymbol{220})} = \frac{\pi V cos\theta_B}{\lambda F_{\boldsymbol{g}}} = \frac{\pi *\ 163.6\text{Å}^3 * 1}{3.70e - 2\text{Å} *\ 15.06\ \text{Å}} = 92\ nm$$
$$\xi_{\boldsymbol{DJ2}|(\boldsymbol{220})} = \frac{\pi V cos\theta_B}{\lambda F_{\boldsymbol{g}}} = \frac{\pi * 2627\text{Å}^3 * 1}{2.97e - 3\text{Å} *\ 261.73\ \text{Å}} = 1062\ nm \tag{S3.8}$$

The calculation of Si is close to the value reported in the literature (75.7 nm)[5], giving a fair estimation. Compared to the sample thickness of ~270nm of DJ n=2 crystal, the extinction distance is ~4 times larger, therefore we believe the multiple scattering effects are not likely to happen in DJ n=2 perovskites.

**Supplementary discussion 4.** Estimation of antiferro-distortion in 2D perovskites

Fig. S5 shows the simulated diffractions with intrinsic (Fig. S5a) and reduced antiferro-distortion (Fig. S5b), as well as the corresponding crystal structures with different disordering parameter viewed along stacking axis. The top (in-plane) view of 2D perovskite caused by intrinsic in-plane cell doubling is shown in the schematics in Fig. S5c. The tilted angle between adjacent octahedra ($2\theta=24^o$) is associated with antiferro-distortion, forming an initially doubled unit cell of ~8.8 Å (indicated in red squares), as opposed to the undistorted structure ($\theta=0^o$) where unit cell is of ~6.2 Å (labeled by green squares). To estimate the influence on crystal structure with reduced antiferro-distortion, we've constructed a hypothetic structural phase (Fig. S5d), where interlayer octahedra are more aligned and simulated an electron diffraction pattern correspondingly (Fig. S5b). For a direct comparison between the symmetrized and the intrinsic antiferro-distorsive phase, a subtracted diffraction pattern is displayed in Fig. 2c. As we can see, the intensity variations are expected in specific Bragg planes, including {400} and {220} (bright red) increasing and {310} and {530} (blue).

To estimate the light-induced rotation angle $\Delta\theta = \theta' - \theta_0$, we calculated the percentage of Bragg peak intensity variation as a function of tilting angle (Fig. S5e), and compare them with the experiment data at early time (t = 2ps). From the responses of {310}{400}, we estimated the optimal distortion parameter to be ~11.75$^o$, as shown in the shadow region in Fig. S5e, corresponding to an octahedral tilt of $\Delta\theta = 0.25\ ^o$

**Supplementary discussion 5.** Transient absorption spectroscopy of DJ n=2 crystal and carrier temperature extraction.

The transient absorption (TA) measurements are performed on nearly identical 2D perovskite crystals (similar lateral size ~200um, thickness of ~300nm and composition (DJ n=2) as used for UED). We used an optical pump set at 3.1 eV (same as UED) with 0.1~0.75 mJ/cm$^2$ (1.1×10$^{12}$ cm$^-$

$^{2}$~8.4×10$^{12}$ cm$^{-2}$, corresponding to an excitation regime similar than in the UED measurements), followed by a white-light probe monitoring the transmission variations of the sample (see Method Section on Transient absorption spectroscopy). Therefore, for similar optical pump conditions, we had access to an optical probe in TA, yielding complementary information with respect to the structural ones provided by the UED experiment. The results are displayed below in Fig. S11.

From our TA data, several major observations are summarized:

A. Strong bleaching of exciton states. Fig. S11 (c) shows a variation of absorbance spectrum at varies fluences. We've observed a significant bleaching of exciton peak at 2.18eV monotonically increasing with excitation densities upto 0.75 mJ/cm$^2$. From our TA values extrapolated to a fluence of 1~2 mJ/cm$^2$, we will expect an almost complete bleaching of the exciton resonance leading essentially to a hot electron-hole pair plasma just after the pump in the UED experiment.
B. Hot-carrier cooling dynamics. We estimated the evolution of effective carrier temperature the by fitting the high-energy tail near the band edge with the Maxwell-Boltzmann distribution Fig. S11 (b), which is a commonly used method to approximate the carrier temperature in semiconductors.[8] As estimated in Fig. 3(c), the lifetime of hot-carriers from sub-ps to ~1 ps. The initial carrier temperature is fitted to be ~3700K for 0.75 mJ/cm$^2$. We extrapolate this to be 4100K for a fluence of 1 mJ/cm$^2$ in the UED experiment.

These TA results suggest that hot carriers may be more dominant at this high excitation densities, where many-body interactions and other non-linear processes could play a crucial role, contradictory to the work of Thouin et al where excitonic regime is mainly discussed.[9] This is also confirmed by the quantitate analysis defining the excitation regime and carrier saturation densities vs temperature which are shown below in discussion 6.

**Supplementary discussion 6.** Estimation of saturation densities in DJ n=2 and RP n=2 crystals

We've estimated the carrier temperature in 2D perovskites to be ~4100K for 1mJ/cm$^2$. (Discussion 5). The next step consists in computing the saturation densities for the excitonic resonances in the DJ n=2 compound. For that purpose, the binding energy and the Bohr radius of the 1S exciton resonance are considered.[10] It is usually considered that the binding energy is reduced in DJ compounds by comparison to RP ones for the same n, but precise experimental data are not available for DJ n=2. We started from the known values available for RP n=2:[11] $E_{,b,RP}$=245meV for the binding energy and $a_{B,RP}$=1.60 nm for the effective Bohr radius related to the diamagnetic shift. Next, the effect of the smaller interlayer spacing is estimated using the effective dielectric model of Ishihara et al[12] $\varepsilon_{eff,RP}$=4.9 vs $\varepsilon_{eff,DJ}$=5.5, leading in turn to a reduction of the binding energy $E_{b,DJ}$=204meV. The BSE model[11] was then used to compute the diamagnetic shift and finally $a_{B,RP}$=1.56 nm. A summary of saturation densities at different carrier temperate is plotted in Fig. 3e.

**Supplementary discussion 7** Estimation of long-time Bragg peak response and light-induced atom RMS displacements $\langle u^2\rangle$

Due to the deviation of Debye-Waller response (Fig. S21) of DJ n=2 crystal, the Bragg peaks intensity responses show competing effects between the conventional Debye-Waller effect and incipient anisotropic lattice ordering:

$$I_{total}(\vec{q}) = [I_0(\vec{q}) + \Delta I(\vec{q}, \Delta\theta)]\ e^{-\frac{1}{3}\Delta\langle u^2\rangle\ q^2} \tag{S7.1}$$

Here $I_0(\vec{q})$ stands for the diffraction intensities at rest. The total Bragg peak intensity response have two contributions: i) an initial ordering process represented in $\Delta I(\vec{q}, \Delta\theta)$, where $\Delta\theta$ is the change of distortion angle (shown in Fig. 2b), which is estimated in Supplementary discussion 4. $\Delta I(\vec{q}, \Delta\theta)$ is simulated based on reduction of andiferro-distortion and also shown in Fig. 2c. ii) classical thermal energy transfer involving all the Bragg peaks, represented in Debye-Waller effect as exponential term $e^{-\frac{1}{3}\Delta\langle u^2\rangle\ q^2}$, where $\langle u^2\rangle$ stands for atomic mean squared (RMS) displacement.

The competing effect of the two mechanisms is also indicated in Fig. S8, where considering only the first contribution (gray dashed line, Fig. S8(b)) will not give a good prediction at long delay time (40-80 ps) especially at higher q, while the combining effect of the two (black solid line) matches better with the experimental response. By comparing the simulated intensity change with the Bragg peak response at long time, the Debye-Waller factor is estimated to be ~1.6e-3 $Å^{-2}$, corresponding to light-induced atom RMS displacement of ~ 0.07 Å at carrier density of $2.5\times10^{13}$ $cm^{-2}$.

**Supplementary discussion 8.** Time evolution of diffraction intensities through the UED run.

Fig. S18 shows the time evolution over ~2 hours during the UED run of DJ n=2 crystal. As shown in Fig. S18a, the total diffraction intensity doesn't show a significant drop over ~2 hours, suggesting the crystal structure remains stable throughout the experiment. Fig. S18b shows the comparison of the electron diffractions between 0min and 120mins, where all the Bragg peaks remain existing, with a slight decrease of overall intensities. Additionally, no rings or new peaks were formed through the experiments (Fig. S18c and Fig. S18d), confirming that no evidence of possible sample degradation or damage is found.

## Supplementary Figures S1 – S20

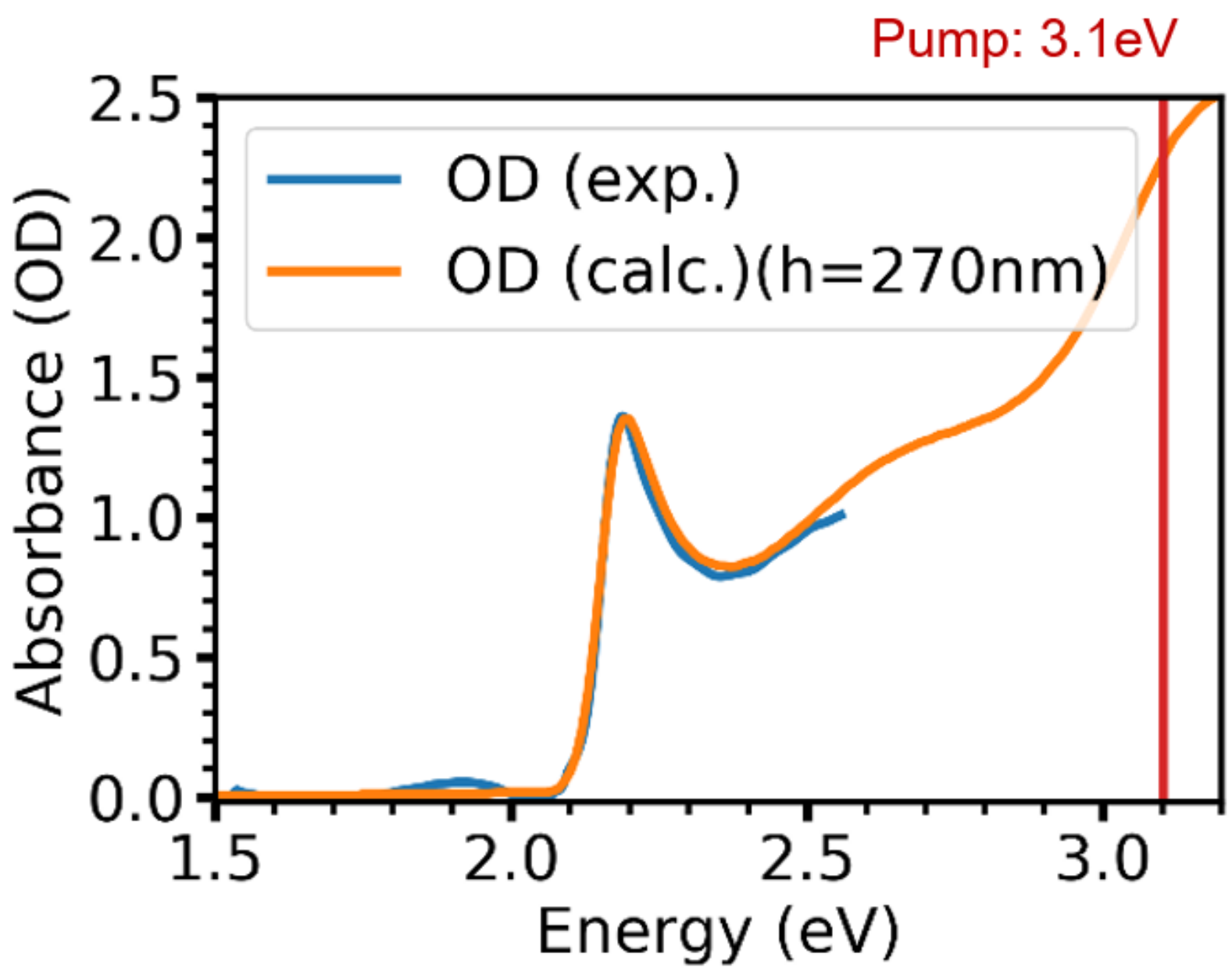

**Fig. S1 | Optical absorbance spectra of DJ n=2 crystal.** The measured absorbance is shown in blue line, with calculated absorbance (orange line) with crystal thickness of 270nm, estimated from the absorption coefficient as reported by Song et al.[3]

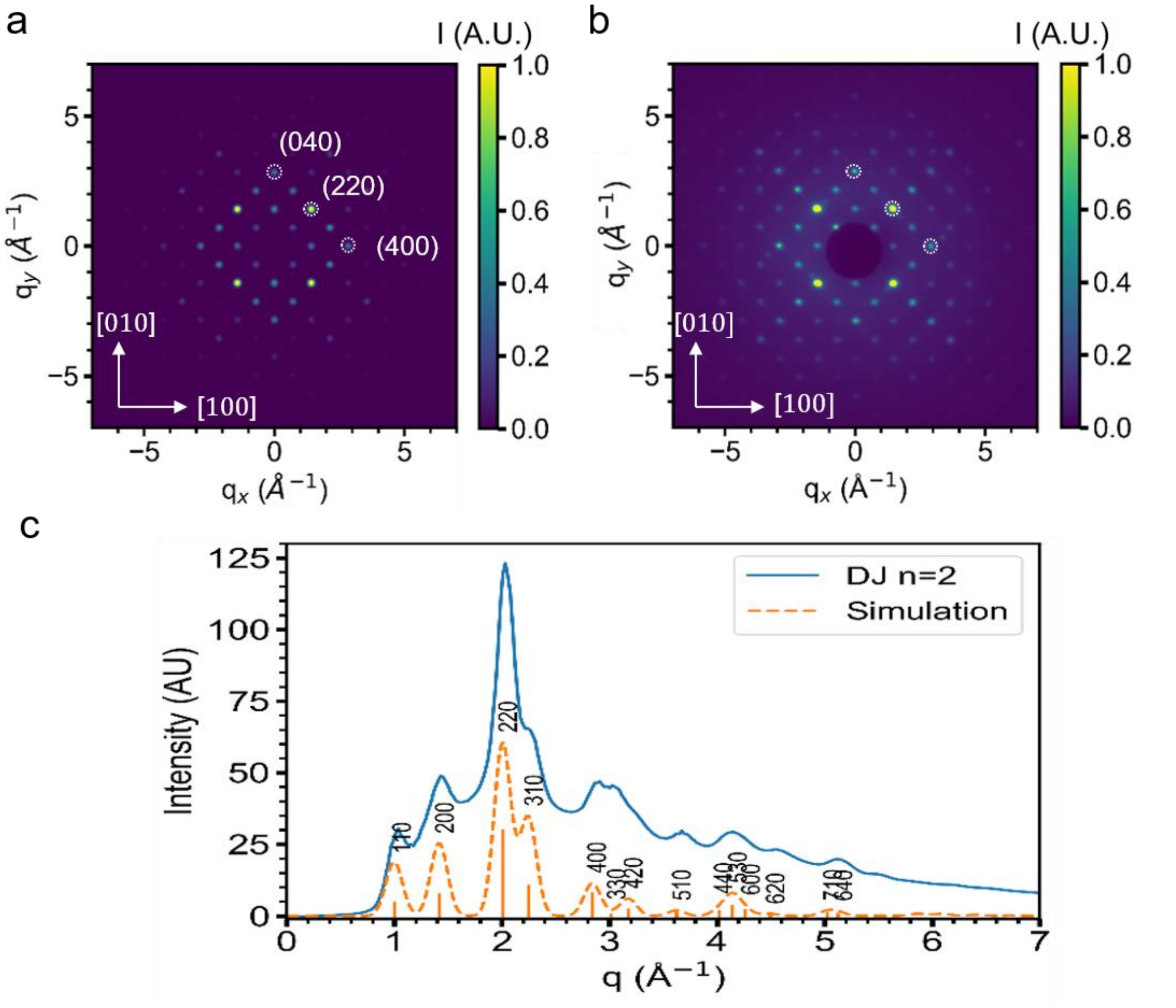

**Fig. S2 | Simulated Diffraction Pattern & Indexing of DJ n=2 crystal. (a)** Simulated Diffraction of DJ n=2 crystal, viewing from [001] direction. **(b)** Static Diffraction of DJ n=2, showing same q-range as the simulated one in Fig. S2a. All the Bragg peaks are well-identified. **(c)** Angular-integrated diffraction of DJ n=2, showing both experimental (blue solid line) and simulated (orange dashed line) plots, as well as Indexing of Bragg peaks (hk0).

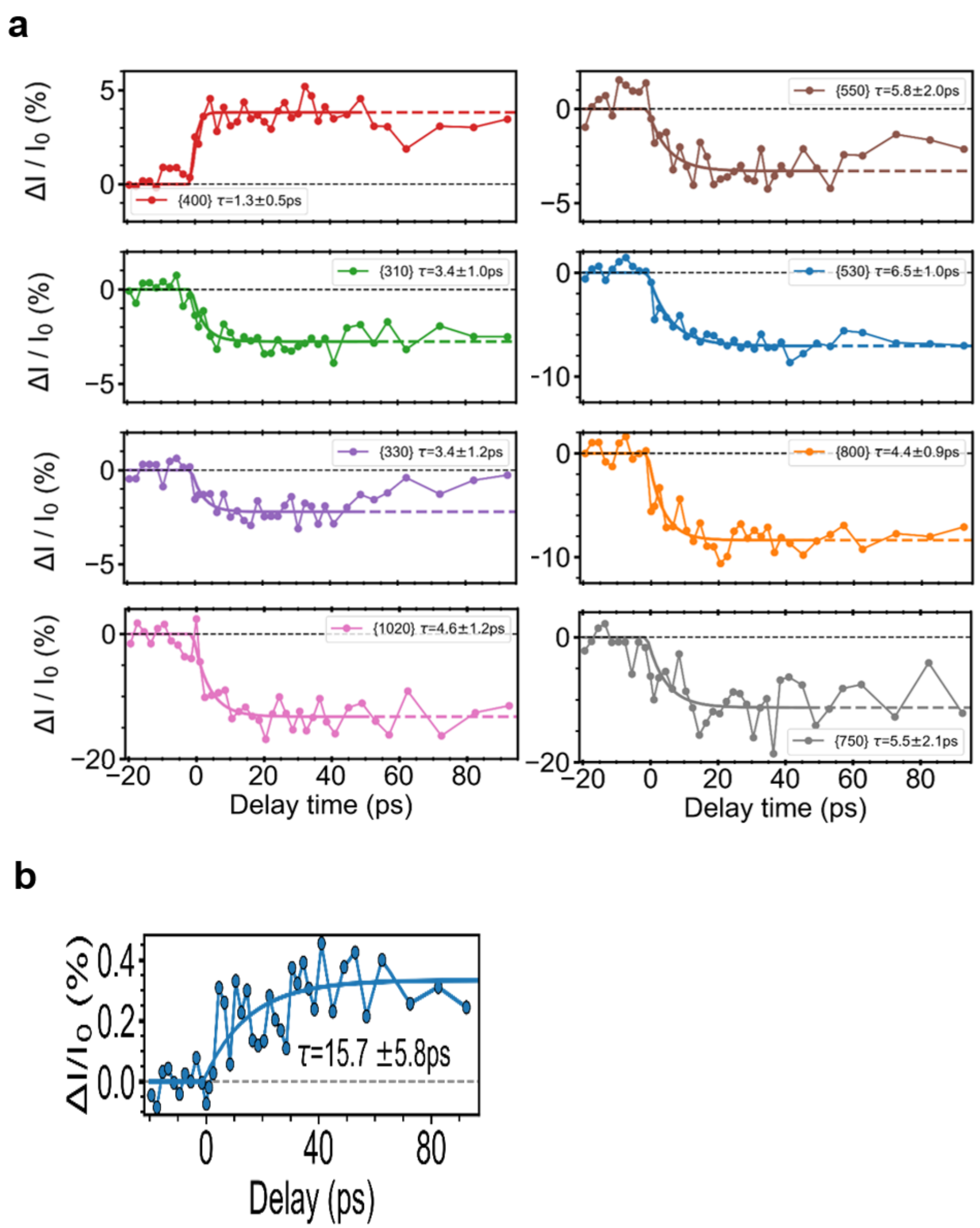

**Fig. S3 | Time traces of selected Bragg peak families in DJ n=2 crystal. (a)** Time traces of Bragg peaks intensities of {400} {310} {330} {1020} {550} {530} {800} and {750} for DJ n=2 crystal. Errors of time constants are estimated by one standard deviation errors from non-linear least-squares fitting. **(b)** Dynamic plots of the total integrated diffuse scattering intensities between Bragg peaks.

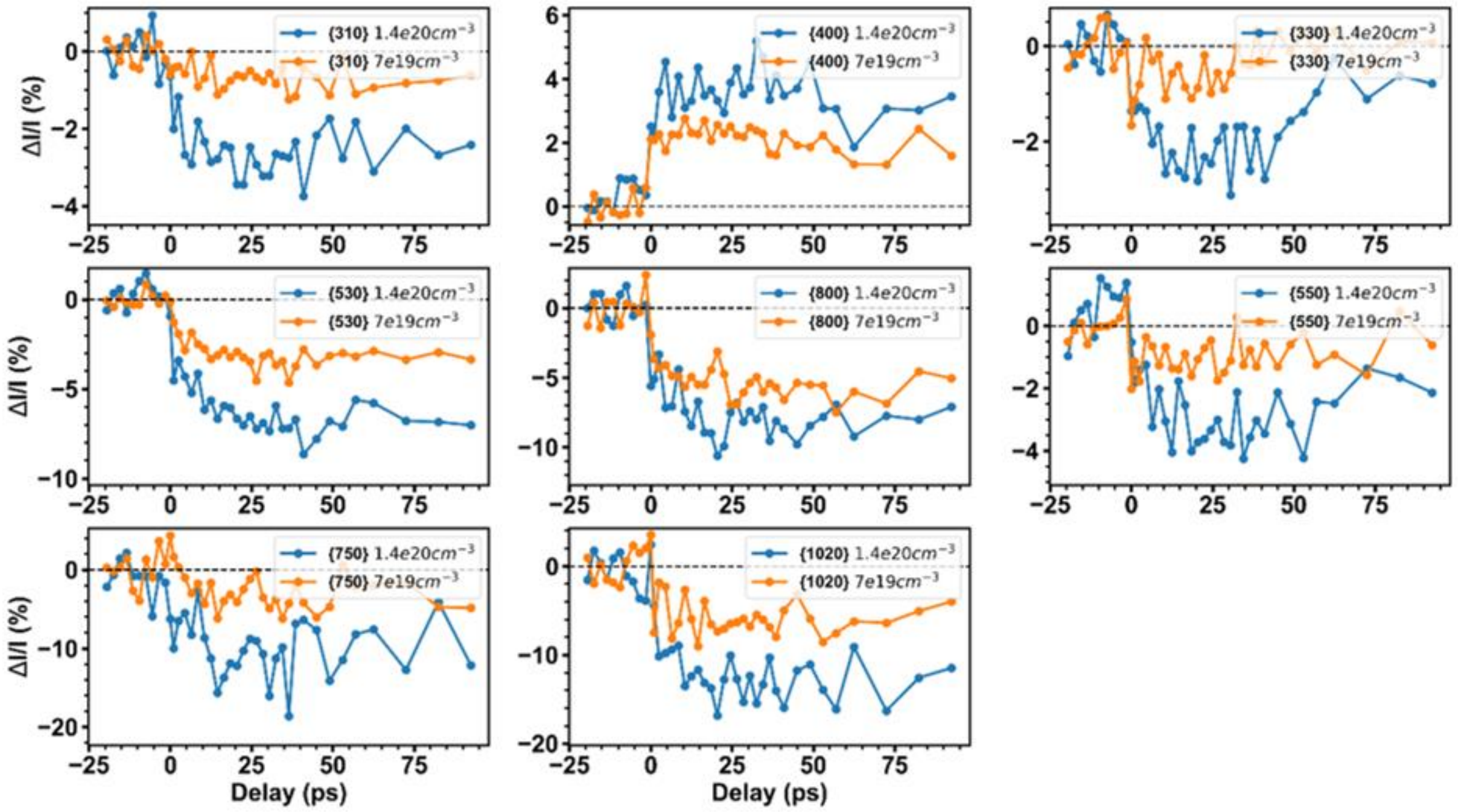

**Fig. S4 | Fluence dependence for selected Bragg peaks in DJ n=2 crystal.**

Time traces of Bragg peaks intensities of {310} {400} {330} {530} {800} {550} {750} and {1020} for DJ n=2 crystal under 2 laser fluences: 1mJ/cm$^2$ (orange) and 2mJ/cm$^2$ (blue).

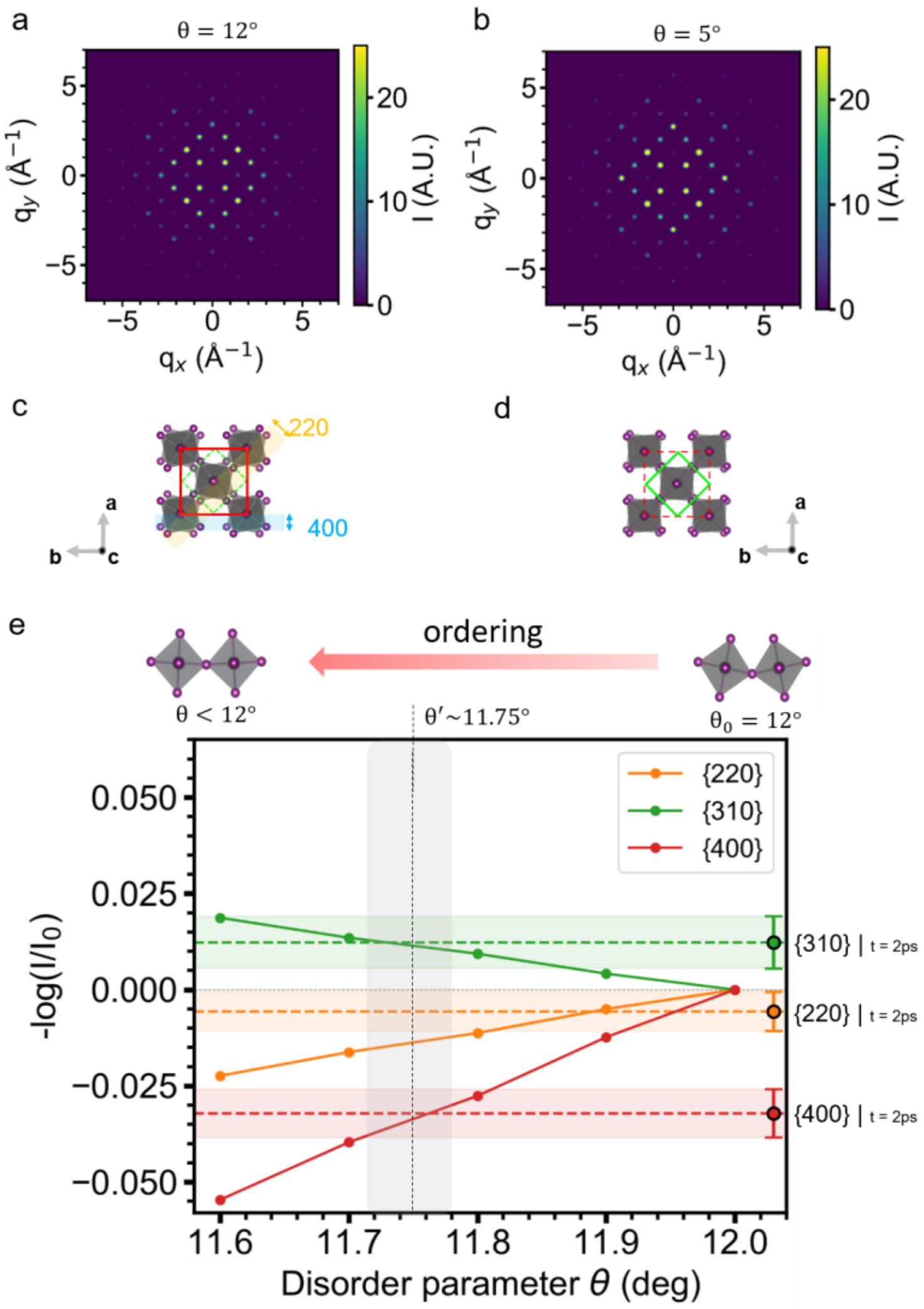

**Fig. S5 | Simulated Diffraction Patterns with tuned antiferro-distortion. (a)** Simulated Diffraction of DJ n=2 at rest with intrinsic antiferro-distortion ($\theta=12^{o}$). **(b)** Simulated Diffraction of DJ n=2 with reduced antiferro-distortion ($\theta=5^{o}$). **(c)(d)** Corresponding schematics of distorted ($\theta=12^{o}$) and less distorted phase ($\theta<12^{o}$), respectively. **(e)** Simulated logarithmic intensity change ($-\log(I/I_0)$) for selected Bragg peaks as a function of disorder parameter θ, shown for {220} {310} and {400} peaks (orange, green, and red solid lines), respectively. Compared with experimental data (DJ n=2, $2.6\times10^{13}$ $cm^{-2}$) at t=2ps (scattered points at right), the optimal change of disorder parameter is estimated to be $\Delta\theta \sim 0.25^{o}$ (regime indicated in gray shadow).

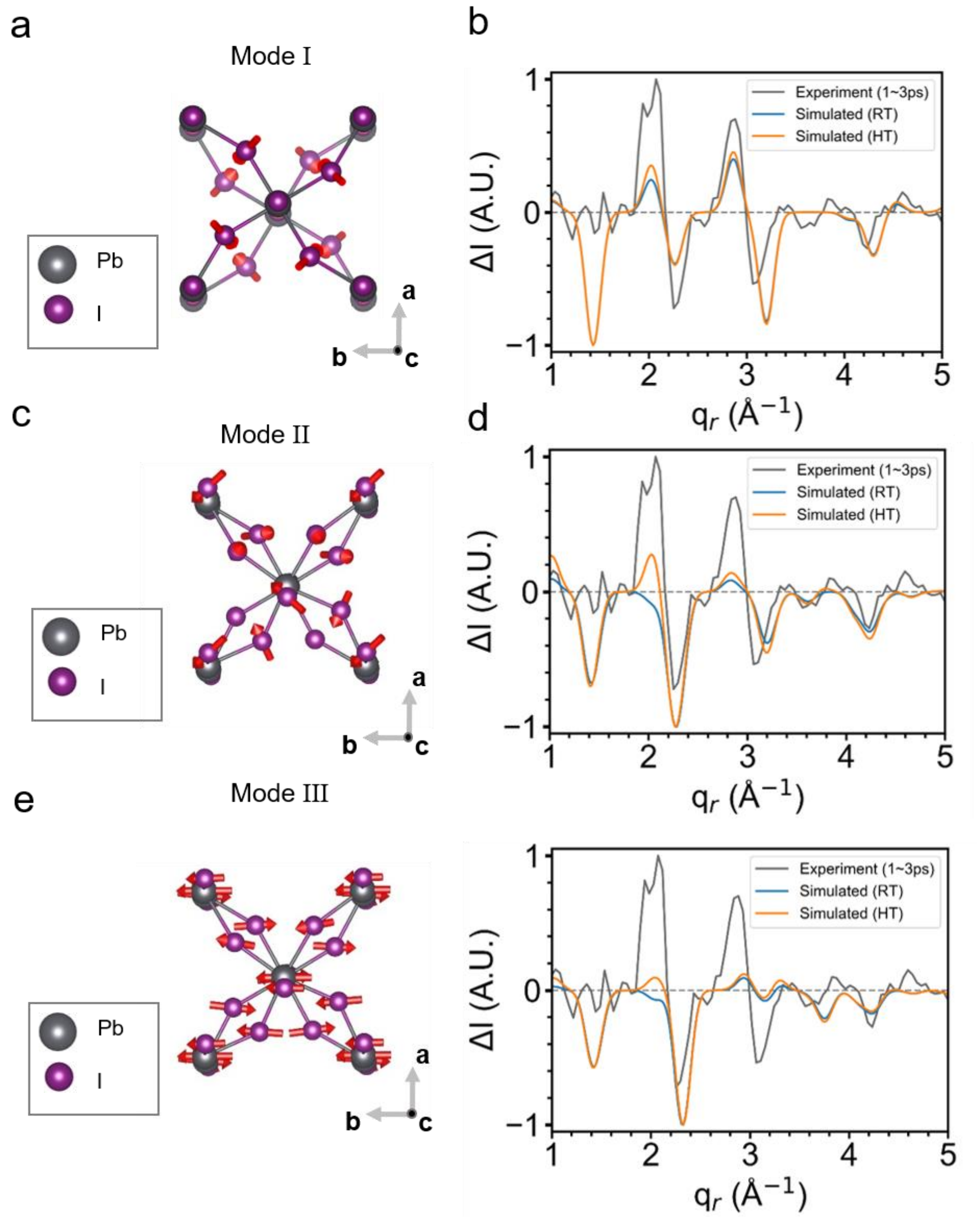

**Fig. S6 | Examples of calculated antiferro-distortive phonon modes candidates in DJ n=2 crystal. (a)(c)(e)** Schematic illumination of calculated normal modes (I, II and III respectively) acquired from phonon simulation (cations are neglected for visualization). The directions of the atomic displacements are indicated with red arrows, suggesting a modulation of antiferro-distortions. **(d)(e)(f)** Simulated changes of electron diffraction plots for normal mode I, II and III respectively, estimated under room temperature (RT, blue solid line) and high temperature (HT, orange solid line) regimes. These simulated plots are compared with differential diffraction plot from UED experiment (1~3 ps, gray solid line), showing consistent peak increase at (220) and (400).

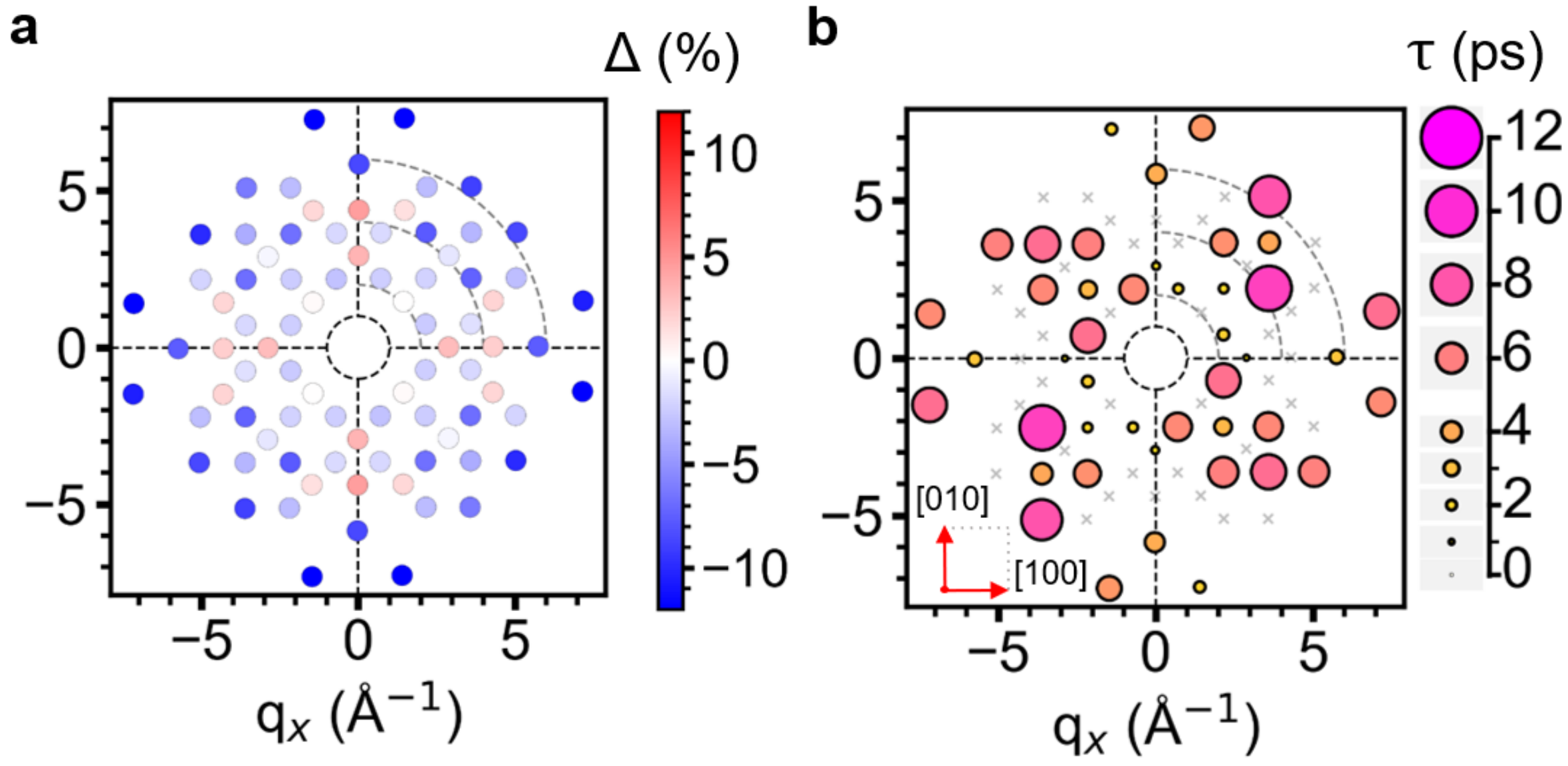

**Fig. S7 | Scatter plots of fitted intensity change and time constants of DJ n=2 crystal. (a)** Scatter plots of fitted intensity change (Δ (%)) of the Bragg peaks (hk0). Each Bragg peak (hk0) is averaged with (-h-k0) based on point symmetry. **(b)** Scatter plots of fitted time constant of the Bragg peaks, represented in both color scale and circle size (small and orange circles for short time, and large and pink circles for long time). 'x' markers stand for the Bragg peaks that cannot be well fit.

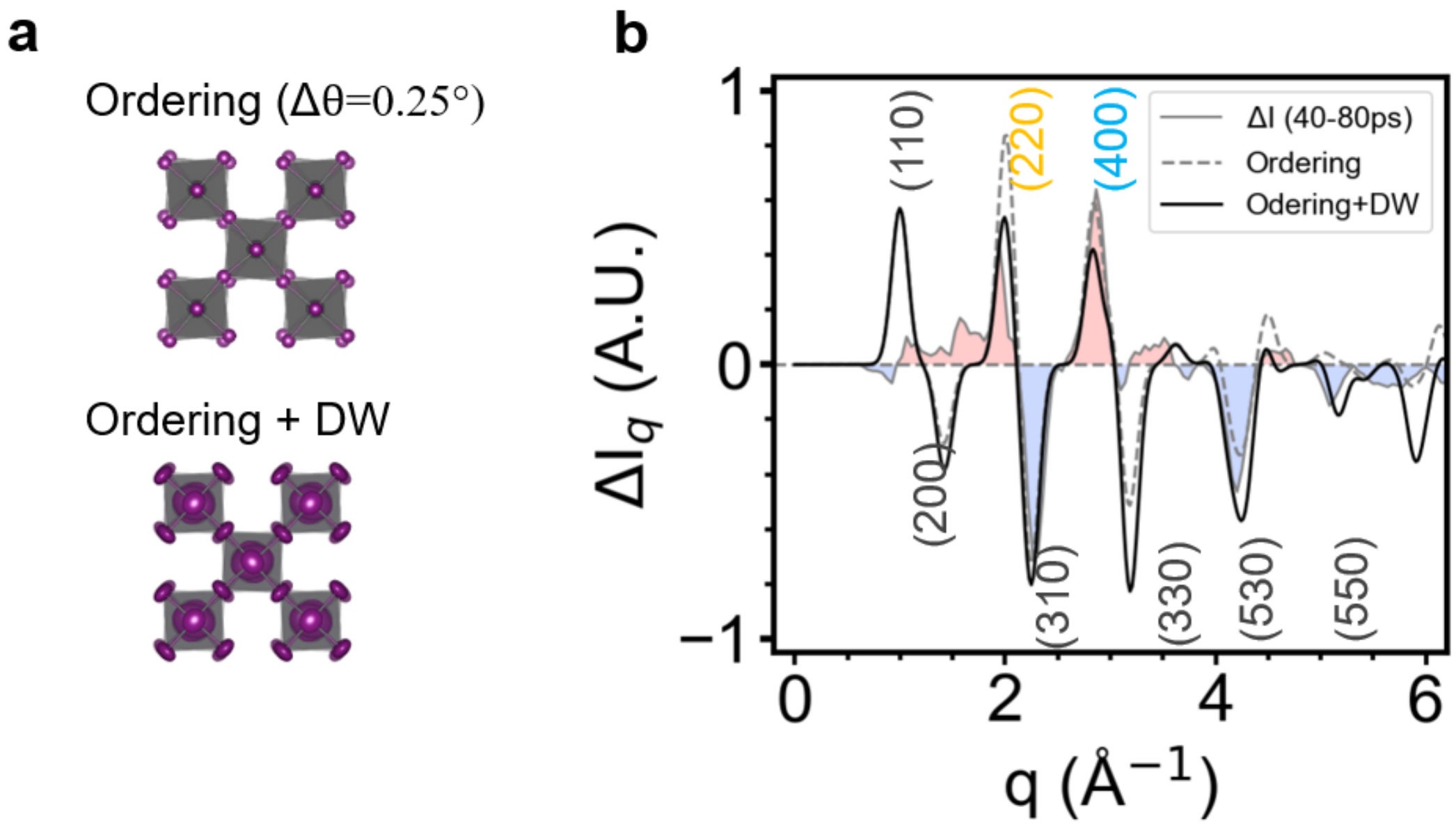

**Fig. S8 | Simulation of total UED response of DJ n=2 crystal. (a)** Simulated crystal structure (top view) for different simulation methods: lattice ordering by reducing distortion angle (top), and additional disorder involving all atoms (bottom). **(b)** Angular integrated differential diffraction plot, averaged between 40-80 ps (gray solid line and shadow), compared with the simulated response with only considering lattice ordering (gray dashed) and both ordering and Debye-Waller effect (black solid line). The experimental intensity around the (110) peak is partially blocked.

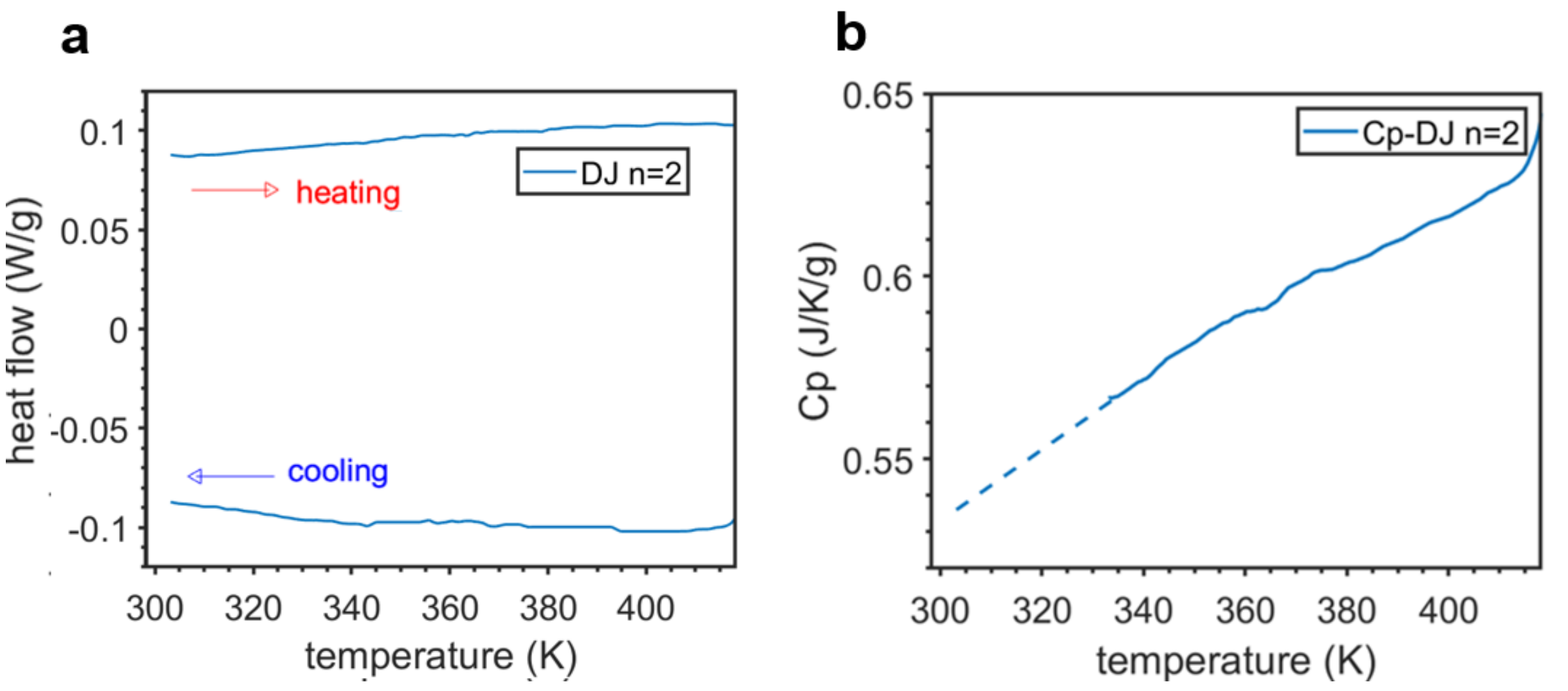

**Fig. S9 | Differential scanning calorimetry (DSC) of DJ n=2 crystal. (a)** DSC curve of DJ n=2 single crystals from room temperature to 150℃. **(b)** Calculated specific heat capacity (Cp) from DSC.

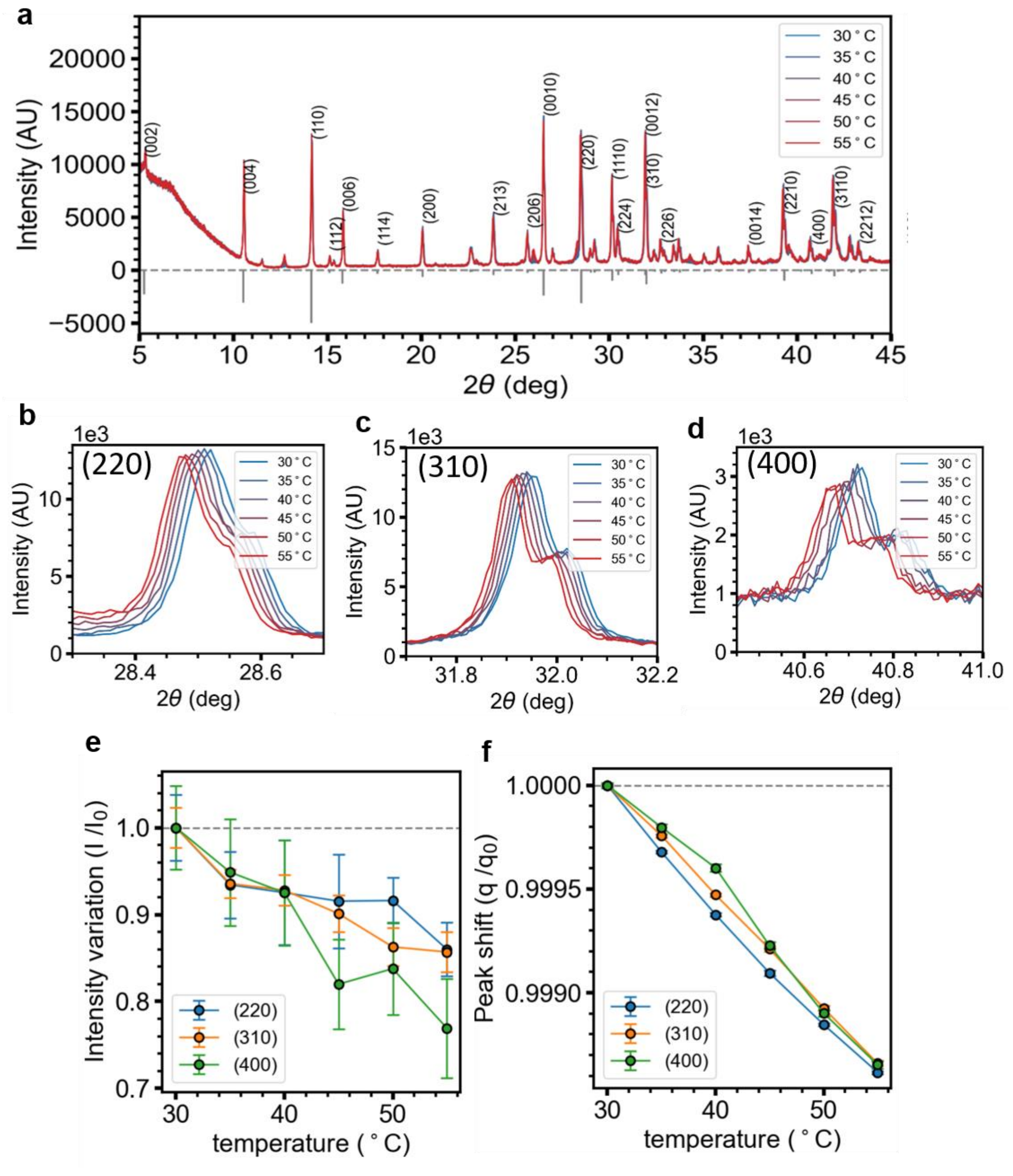

**Fig. S10 | Temperature dependent XRD of DJ n=2 crystal. (a)** Temperature dependent XRD from 30°C to 55°C. Simulated Bragg peak locations for DJ n=2 are indicated in vertical gray solid lines. **(b)(c)(d)** Zoomed-in XRD plots at in-plane peaks (220) (310) and (400) respectively, showing global shifting of peak locations (thermal expansion) and slight intensity decreases over temperature. **(e)** Analysis of relative intensity variation ($I/I_0$) and **(f)** relative peaks shift ($q/q_0$) for (220), (310) and (400) Bragg peaks.

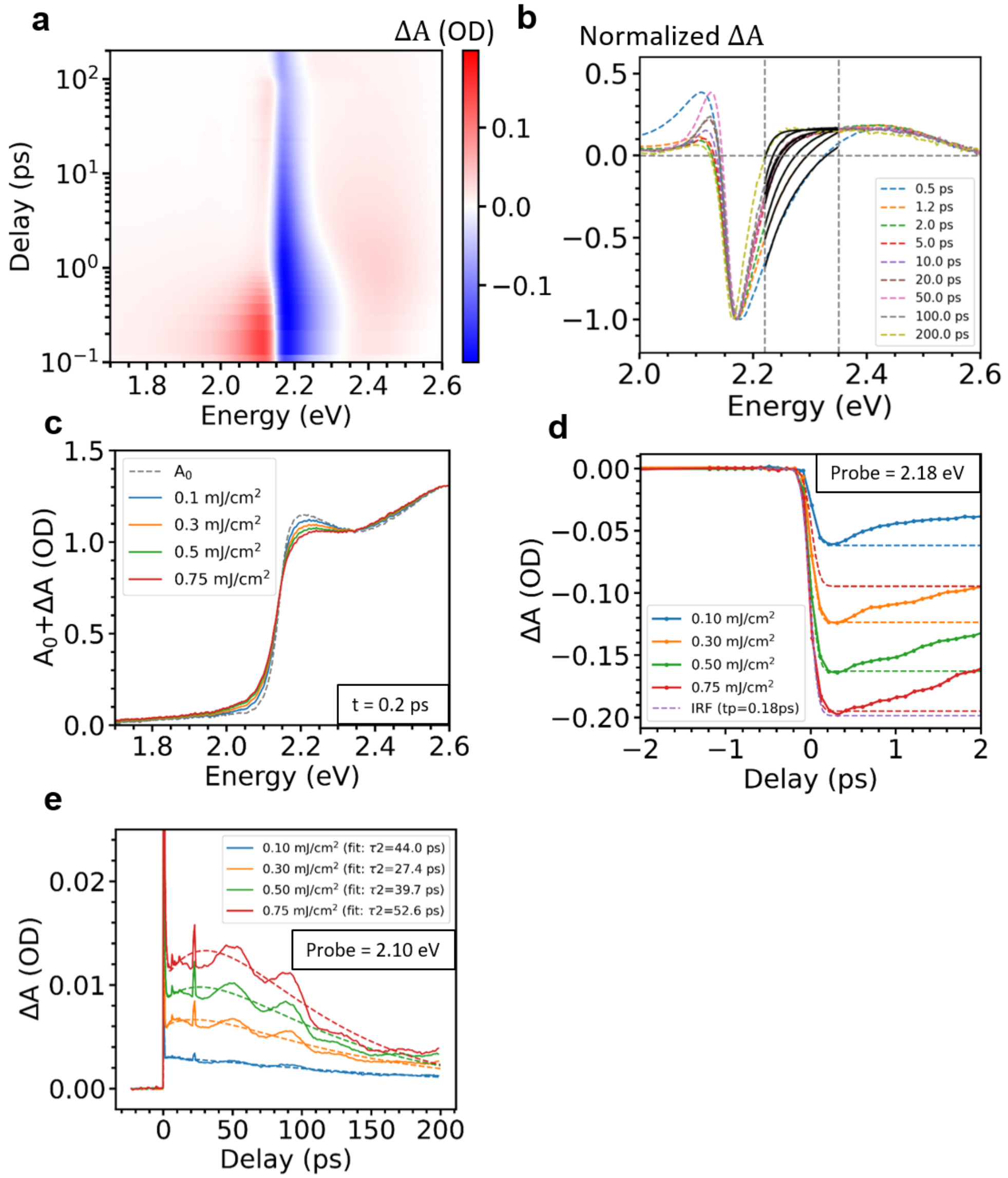

**Fig. S11 | Transient absorption spectroscopy for DJ n=2 crystal. (a)** Pseudo color representation of transient absorbance (Δ OD) as function of delay time, under 0.75mJ/cm$^2$. **(b)** Normalized TA spectra. The bleach amplitude at 2.18 eV is normalized to -1. High energy tails are fitted by Boltzmann distribution (black solid lines) to extract carrier temperature (shown in Fig. 3c). **(c)** Absolute absorption at various fluences, showing almost complete bleaching of exciton resonance. **(d)** Ultrafast bleaching ($\tau \leq 0.18$ ps) of exciton resonance at 2.18 eV, indicating fast ground state exciton bleaching after above-band excitation. **(e)** Observation of coherent acoustic phonon oscillations for DJ n=2 in the DW regime. The period of oscillations is roughly estimated to 45ps.

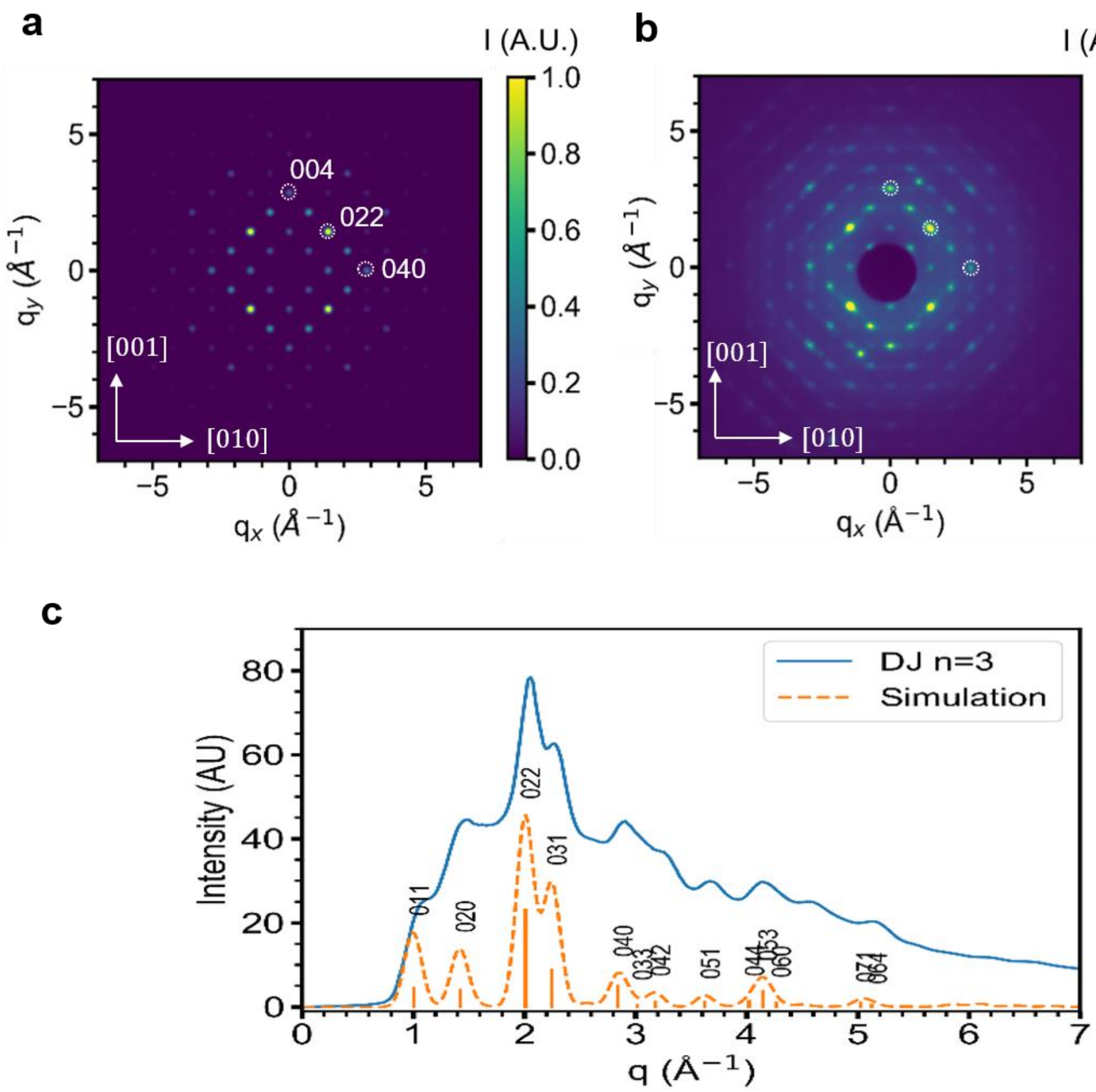

**Fig. S12 | Simulated Diffraction Pattern & Indexing of DJ n=3 crystal.**

**(a)** Simulated Diffraction of DJ n=3, viewing from [100] direction. **(b)** Static Diffraction of DJ n=3, showing same q-range as the simulated one in Fig. S12a. All the Bragg peaks are well-identified. **(c)** Angular-integrated diffraction of DJ n=3, showing both experimental (blue solid line) and simulated (orange dashed line) plots, as well as Indexing of Bragg peaks (0hk).

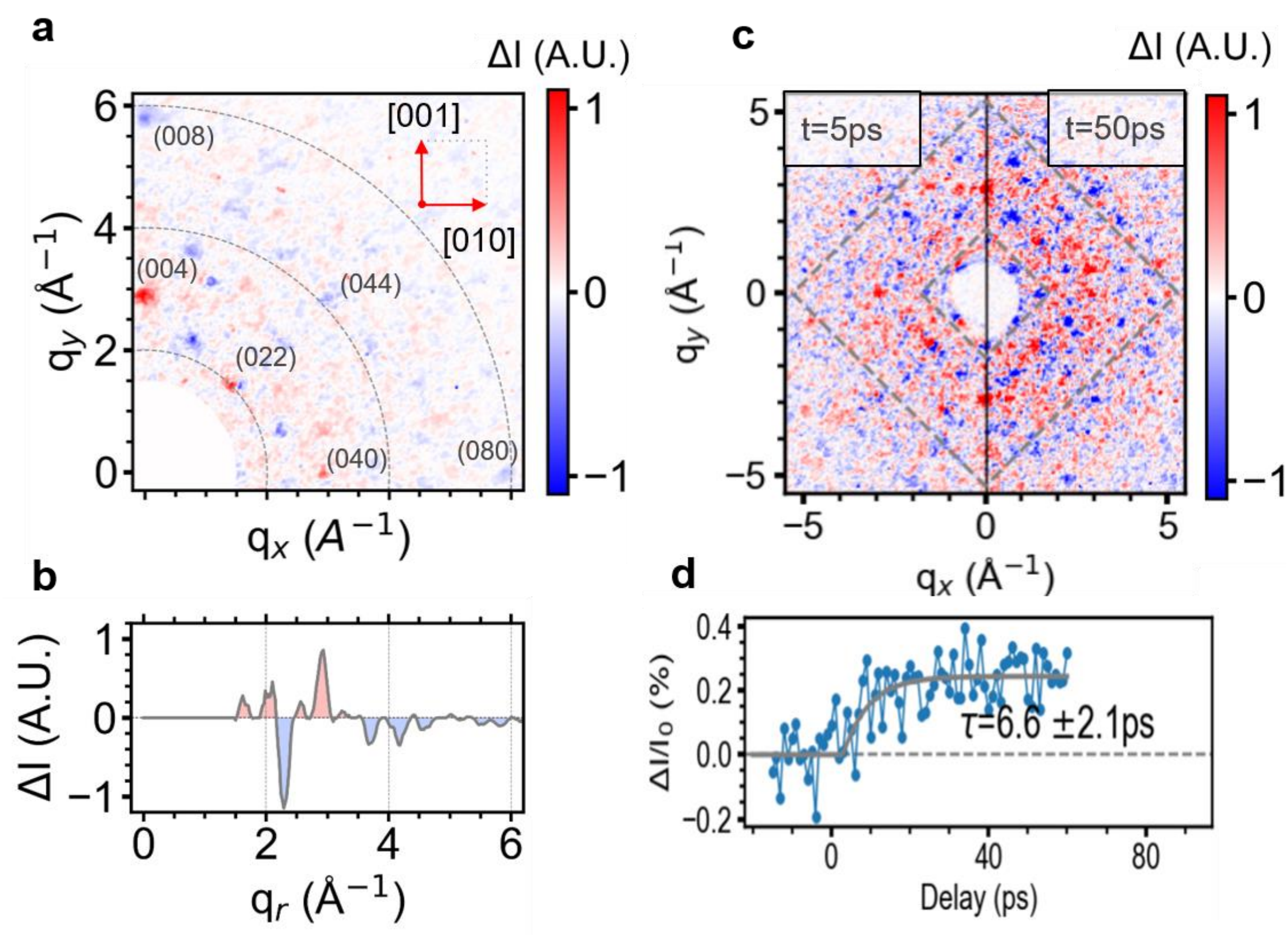

**Fig. S13 | Differential diffraction patterns of DJ n=3. (a)** Differential diffraction pattern (20~40ps) for DJ n=3, showing similar response as DJ n=2, including the positively-changing peaks (040), (022) and (004). **(b)** Angular-integrated differential diffraction (20~40ps) for DJ n=3. **(c)** Differential diffraction pattern at 5ps (left) and 50ps (right). The increasing red colors in between the peaks correlate to the diffuse scattering intensities. **(d)** Time traces of the integrated diffuse scattering intensities integrated with the two squares in **(c).**

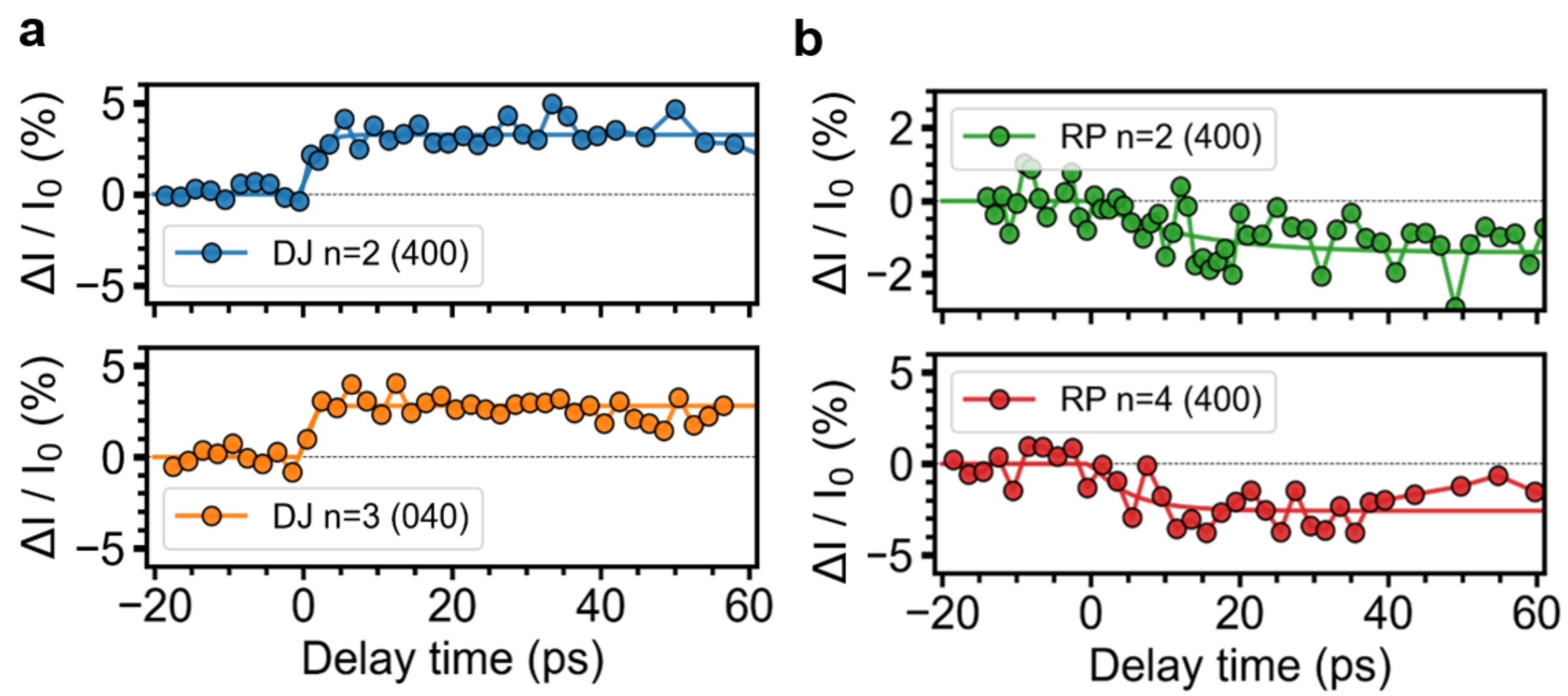

**Fig. S14 | Comparison of time traces of selected Bragg peaks in 2D perovskites. (a)** Time traces of DJ n=2 (400) (blue) and DJ n=3 (040) (orange) peak. Both peaks show a rapid increase in intensity. The rise-time constant fitted to be ~1.3 ps (DJ n=2, blue solid line) and ~0.9 ps (DJ n=3, orange solid line) respectively. **(b)** Time traces of RP n=2 (400) (green) and RP n=4 (400) (red) peak. Both peaks show a slower increase in intensity variations.

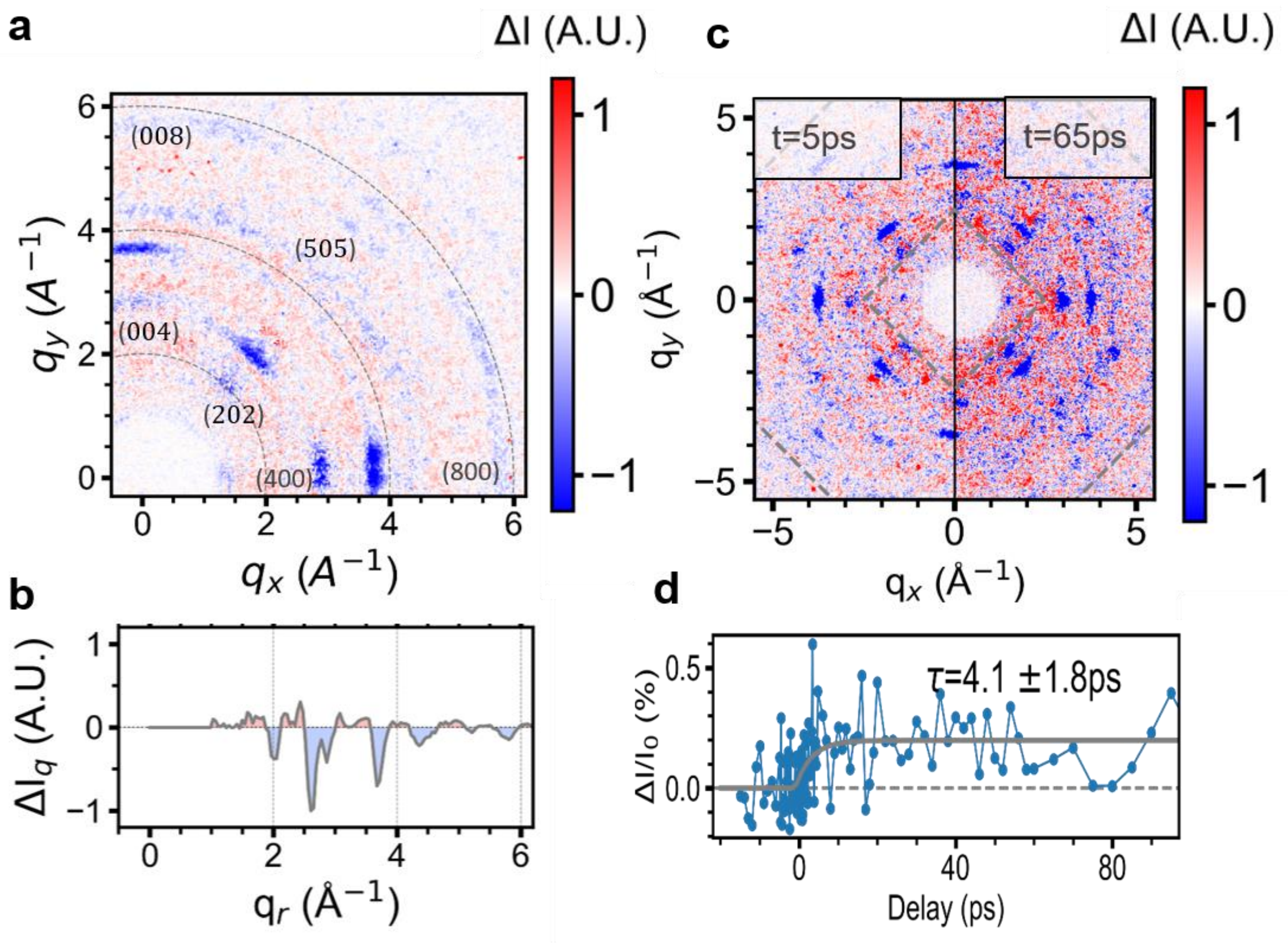

**Fig. S15 | Differential diffraction patterns of RP n=2. (a)** Differential diffraction pattern (15~40ps) for RP n=2, showing different response than DJ samples where no peaks increasing observed. **(b)** Angular-integrated differential diffraction (15~40ps) for RP n=2. **(c)** Differential diffraction pattern at 2ps (left) and 65ps (right). The increasing red colors in between the peaks correlate to the diffuse scattering intensities. **(d)** Time traces of the integrated diffuse scattering intensities integrated in-between the two squares in **(c)** (dashed line).

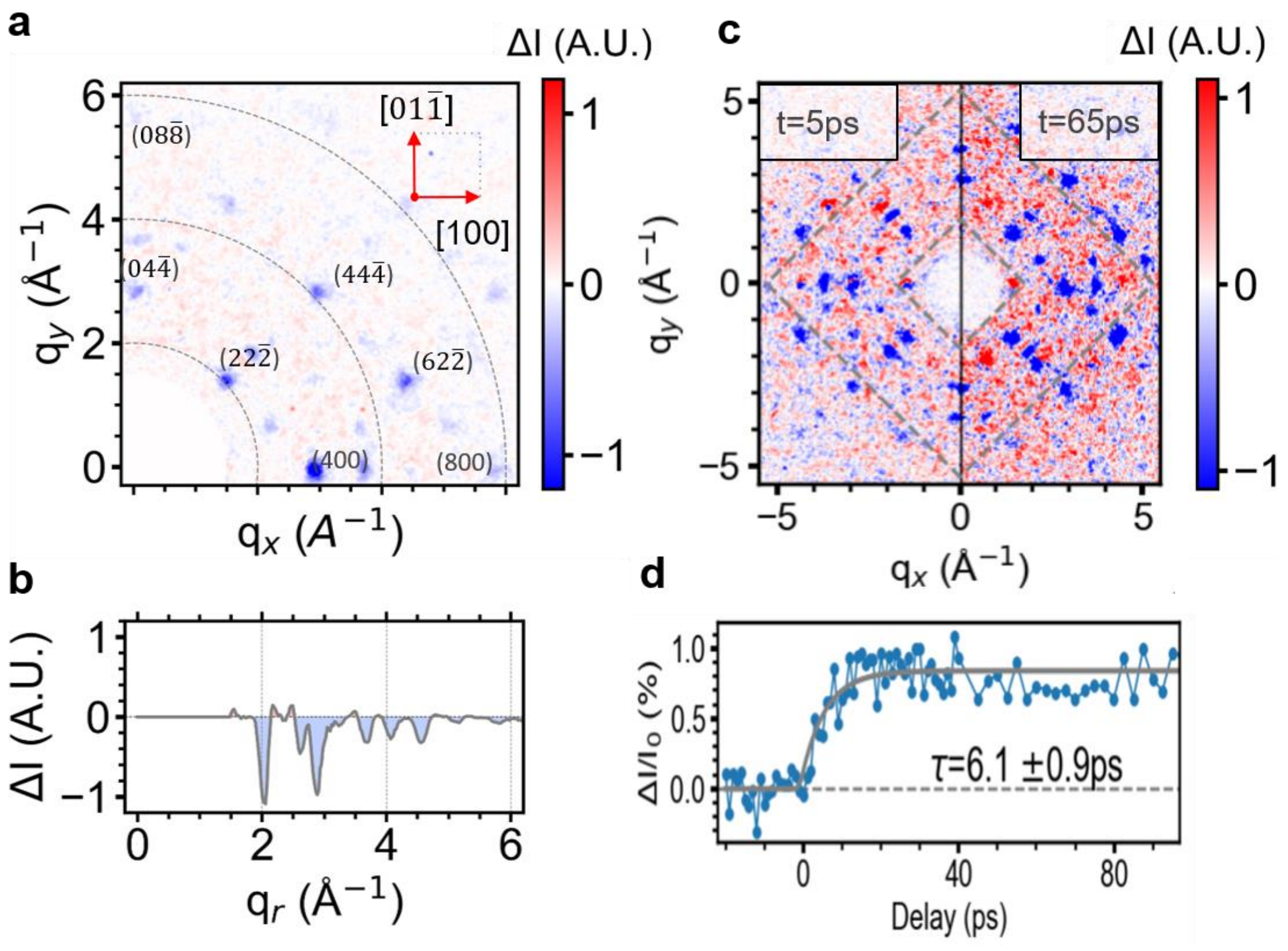

**Fig. S16 | Differential diffraction patterns of RP n=4. (a)** Differential diffraction pattern (20~40ps) for RP n=4, showing different response than DJ samples where no peaks increasing observed. **(b)** Angular-integrated differential diffraction (20~40ps) for RP n=4. **(c)** Differential diffraction pattern at 5ps (left) and 65ps (right). The increasing red colors in between the peaks correlate to the diffuse scattering intensities. **(d)** Time traces of the integrated diffuse scattering intensities integrated with the two squares in **(c).**

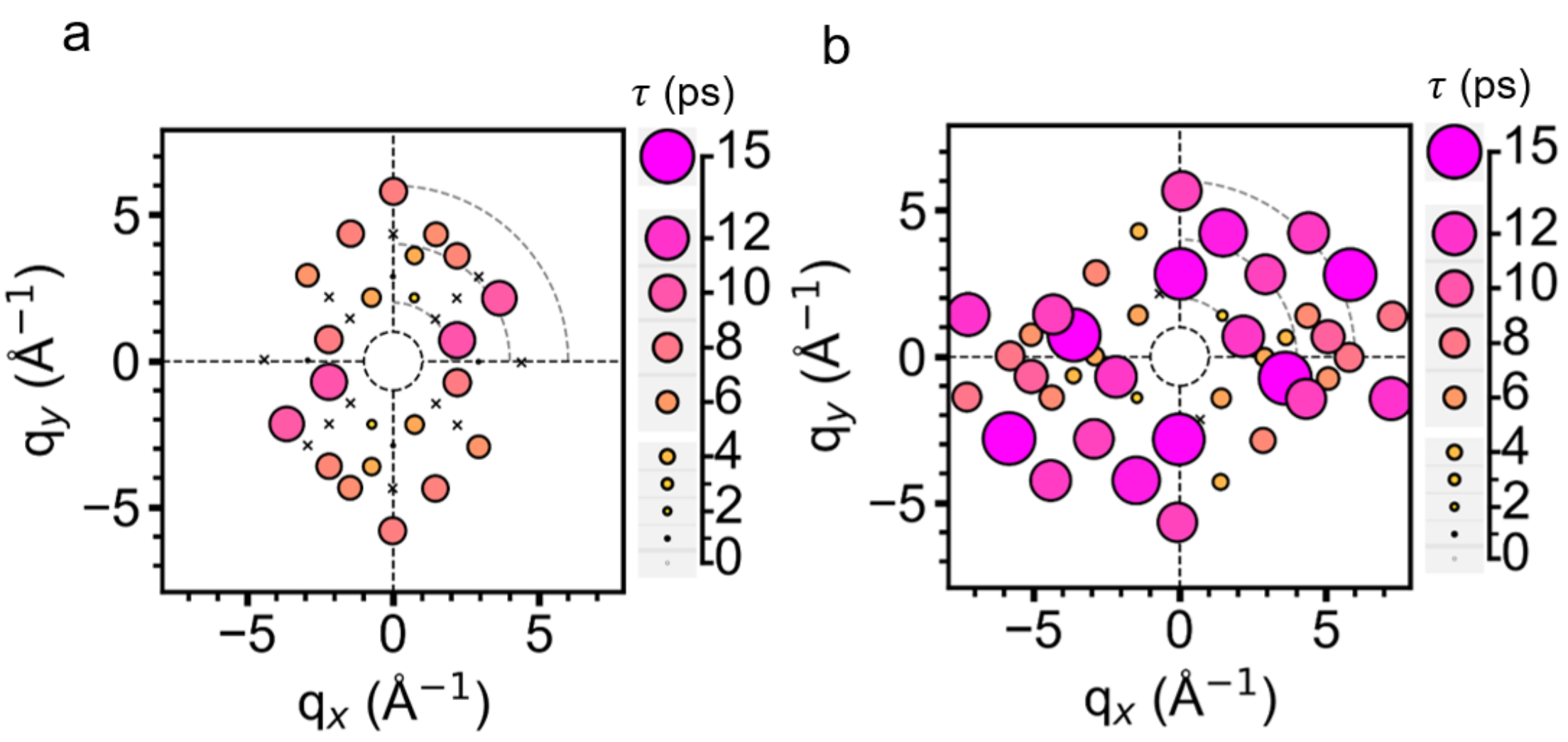

**Fig. S17 | Time response comparison of different samples.** Scatter plots of time constants ($\tau$) of **(a)** DJ n=3 and **(b)** RP n=4 are shown. All the time constants are extracted from single exponential time fitting of Bragg peaks (averaged within each pair (hk0) and (-h-k 0)).

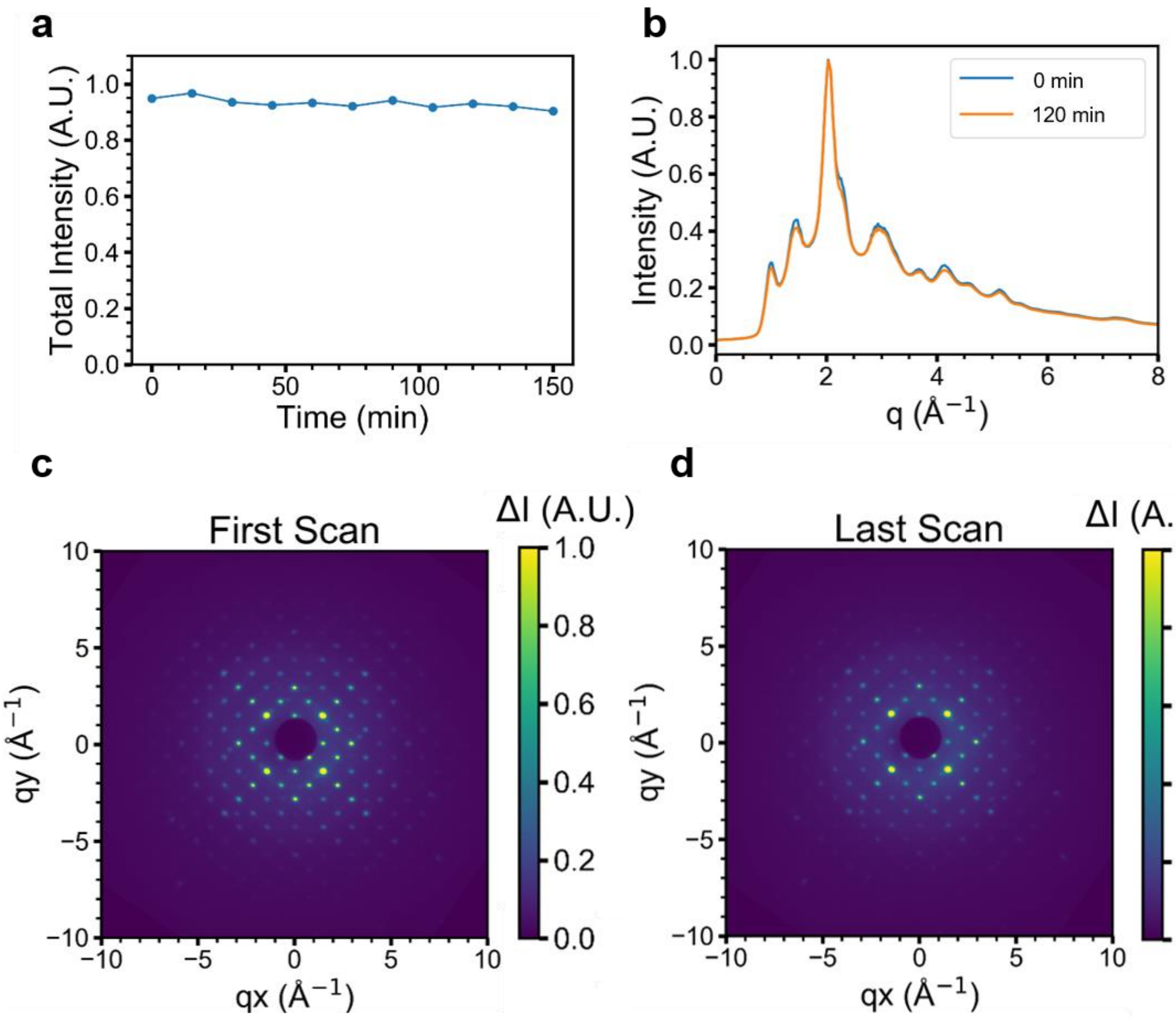

**Fig. S18 | Time Evolution of diffraction intensity of DJ n=2 crystal. (a)** Total electron diffraction intensity during UED measurements over 150mins. **(b)** Azimuth integrated plot of diffraction patterns taken at 0 min (blue) and 120mins (orange) after during the UED run. **(c)(d)** Diffraction images of **(c)** first and **(d)** last scan.

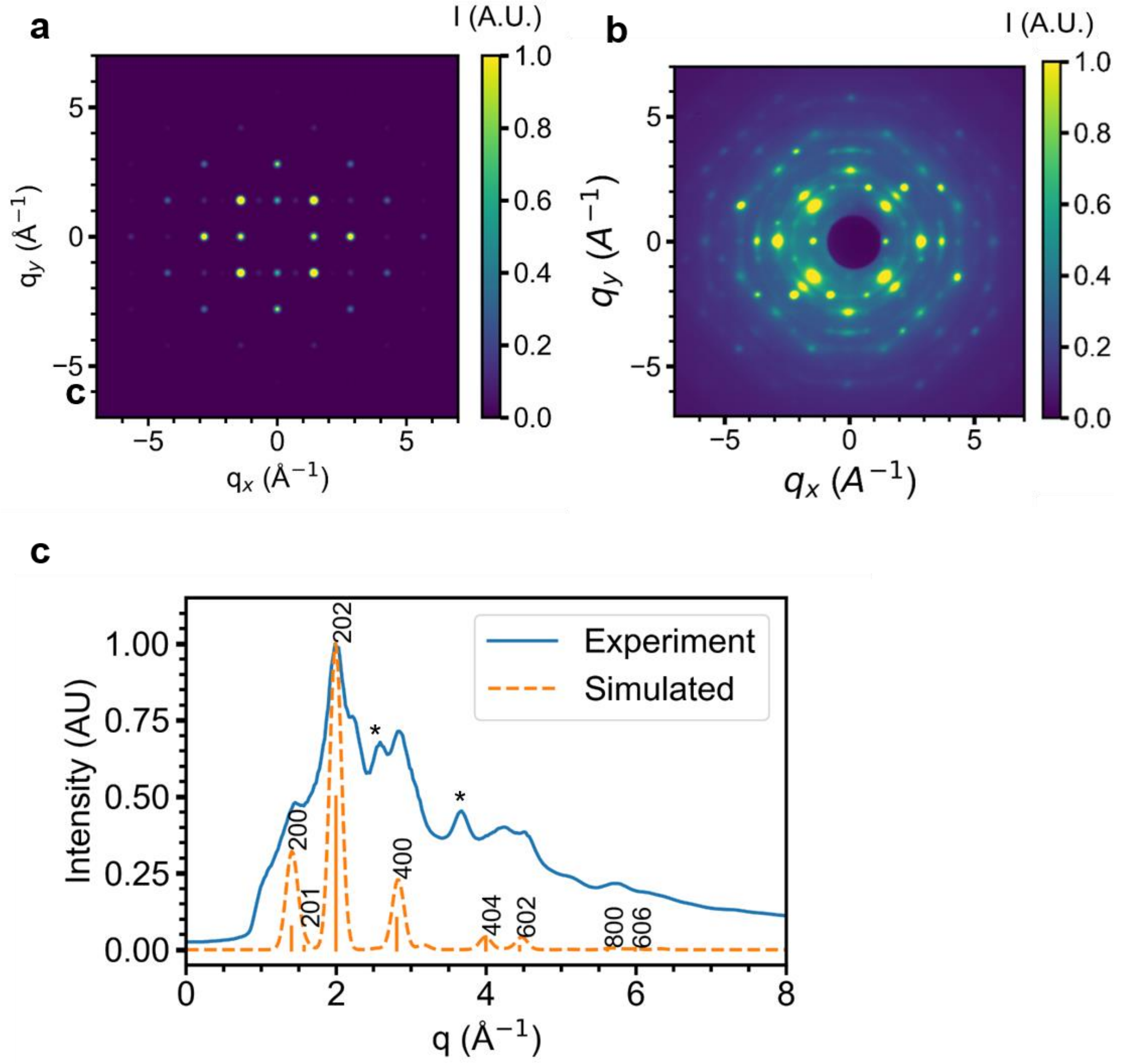

**Fig. S19 | Simulated Diffraction Pattern & Indexing of RP n=2 crystal.**

**(a)** Simulated Diffraction of DJ n=2, viewing from [010] direction. **(b)** Static Diffraction of RP n=2, showing same q-range as the simulated one in Fig. S19a. **(c)** Angular-integrated diffraction of RP n=2, showing both experimental (blue solid line) and simulated (orange dashed line) plots, as well as Indexing of Bragg peaks (0hk).

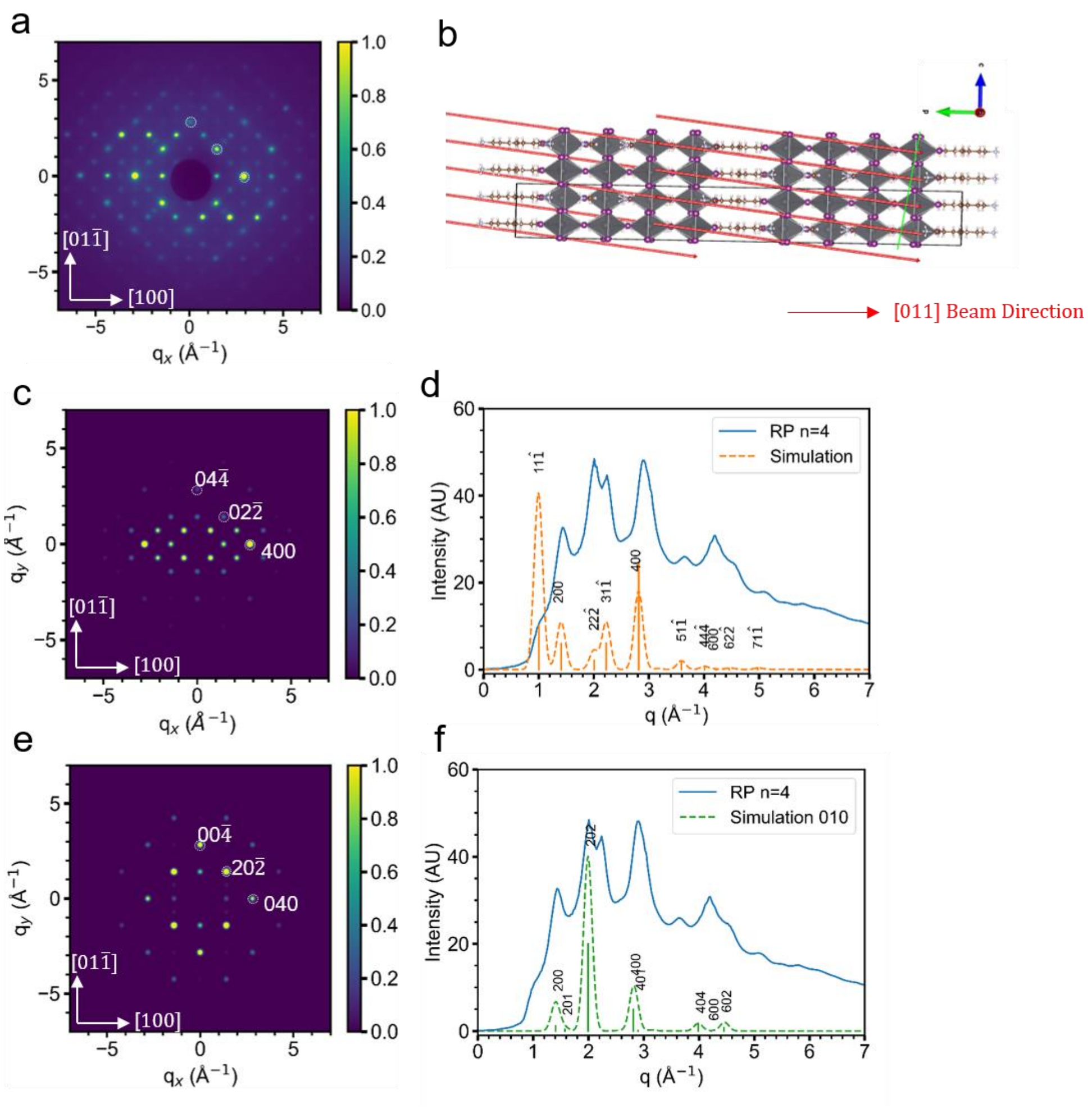

**Fig. S20 | Simulated Diffraction Pattern & Indexing of RP n=4 crystal. (a)** Static Diffraction patterns of RP n=4. **(b)** Schematics of RP n=4 and indication of [011] direction (red arrows). **(c)** Simulated Diffraction of RP n=4 along [011] direction. **(d)** Angular-integrated diffraction of RP n=4 viewing at [011], showing both experimental (blue solid line) and simulated (orange dashed line) plots, as well as Indexing of Bragg peaks. **(e)** Simulated Diffraction of RP n=4 along [010] direction, showing a mismatch with experimental patterns. **(f)** Angular-integrated diffraction of RP n=4 viewing at [010], showing both experimental (blue solid line) and simulated (green dashed line) plots, as well as Indexing of Bragg peaks.

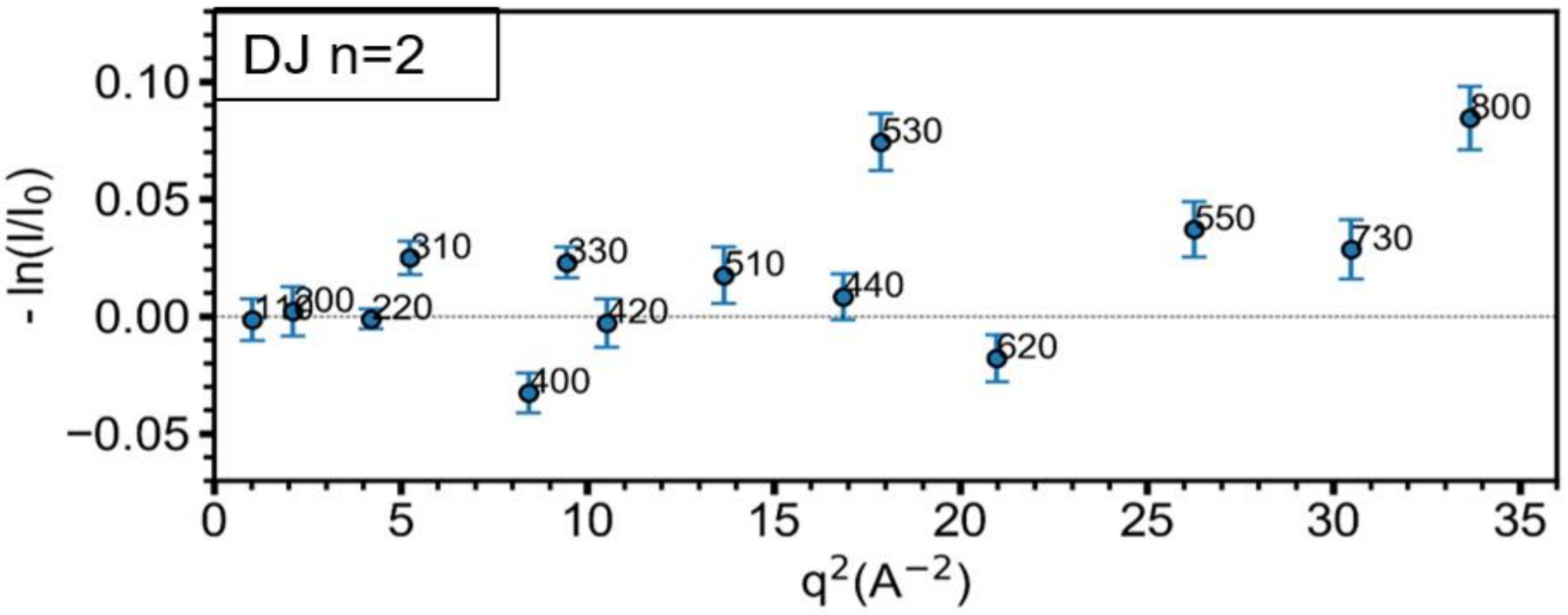

**Fig. S21 | Logarithmic plot of Bragg peaks intensity as a function of $q^2$ for DJ n=2.** Logarithmic plot of Bragg peaks total relative intensity change ($-\log(I/I_0)$) DJ n=2 under pump fluence energy of 2mJ/cm$^2$.

## Supplementary Tables

| Sample Name | DJ n=2 | DJ n=3 | RP n=2 | RP n=4 |
|---|---|---|---|---|
| | | | | |
| Excitation Energy (eV) | 3.1 | 3.1 | 3.1 | 3.1 |
| Fluence ($mJ/cm^2$): | 2 | 2.2 | 0.45 | 2.5 |
| Crystal thickness (nm): | 270 | 450 | 200 | 380 |
| OD (at 3.1eV): | 2.2 | 2.5 | 2.5 | 2.5 |
| Carrier density ($cm^{-2}$) | $2.5\times10^{13}$ | $2.3\times10^{13}$ | $8.9\times10^{12}$ | $4.2\times10^{13}$ |

**Table. S1: Lists of crystal thickness, fluences and carrier densities.** Sample thickness are estimated from the optical absorbance data in Fig. S1 for DJ n=2.